\documentclass[screen,authorversion,nonacm]{acmart}

\usepackage{subfig}
\usepackage{gensymb}
\usepackage{wrapfig}

\begin{document}

\title{Measuring Latency Reduction and the Digital Divide of Cloud Edge Datacenters}

\author{Noah Martin}
\affiliation{%
   \institution{Tufts University}
   \city{Boston}
   \state{MA}
   \country{USA}}
\email{noah.martin@tufts.edu}

\author{Fahad Dogar}
\affiliation{%
   \institution{Tufts University}
   \city{Boston}
   \state{MA}
   \country{USA}}
\email{fahad@cs.tufts.edu}

\begin{abstract}
Cloud providers are highly incentivized to reduce latency. One way they do this is by locating datacenters as close to users as possible. These ``cloud edge'' datacenters are placed in metropolitan areas and enable edge computing for residents of these cities. Therefore, which cities are selected to host edge datacenters determines who has the fastest access to applications requiring edge compute -- creating a digital divide between those closest and furthest from a datacenter. In this study we measure latency to the current and predicted cloud edge datacenters of three major cloud providers around the world. Our measurements use the RIPE Atlas platform targeting cloud regions, AWS Local Zones, and network optimization services that minimize the path to the cloud edge. An analysis of the digital divide shows rising inequality as the relative difference between users closest and farthest from cloud compute increases. We also find this inequality unfairly affects lower income census tracts in the US. This result is extended globally using remotely sensed night time lights as a proxy for wealth. Finally, we demonstrate that low earth orbit satellite internet can help to close this digital divide and provide more fair access to the cloud edge.
\end{abstract}




\maketitle

\section{Introduction}
Latency reduction is a top priority for internet applications. Amazon’s frequently-referenced experiment~\cite{amazon_sales} demonstrated a 1\% loss in sales for every 100ms of increased latency, with similar trends observed in other experiments as well~\cite{google_latency,akamai_latency}. Not only does reducing latency lead to more profit for service providers, but it also enables new, emerging applications with strict latency constraints.  For example, head-tracking applications such as virtual reality and 360\degree{} video streaming require a response time of under 20ms to avert motion sickness~\cite{latency_threshold}. These applications are frequently powered by the cloud, which abstracts physical compute to a datacenter that may be physically far from users.

To support low latency for some of these emerging applications, cloud providers have started offering ``edge locations'' which are even closer to the end users compared to traditional datacenters. This typically entails offering compute (i.e., virtual machines - VMs) at the edges of their private cloud network, the closest they can get to end users. Since 2019, Amazon has launched 27 of these edge locations, named Local Zones,  around the world, with many more announced~\cite{local_zone_locations}. Other cloud providers have announced plans for similar services~\cite{azure_edge_zones, google_cloud_edge}. These examples of the ``cloud edge'' are one version of the growing trend towards edge computing~\cite{created_edge_computing} that is increasingly used in industry~\cite{local_zone_couchbase}. This trend is also supported by the rise of serverless computing~\cite{nightcore} to combine the benefits of the edge with the scalability and ease of use of functions as a service~\cite{lambda_edge, cloudflare_workers, akamai_edge}. Finally, this cloud edge presents an opportunity to support many traditional networking solutions, including support for content and service centric networking~\cite{ndn, xia-nsdi, tapa, slack-ccr, catnap}  

While cloud providers are working to reduce latency, there remains a ``digital divide''~\cite{digitalDivide} separating those with quality Internet access from those without. Internet access is known to affect various aspects of a community's socio-economic well-being~\cite{internet_health_outcomes, internet_social_distancing}. It has even been recognized as a key prerequisite for many of the sustainable development goals (SDGs)~\cite{sdg_ict_blog}. 
Prior work has measured the digital divide's effect on network download speed~\cite{characterizingCaliforniaInternetQuality}, but there has been no systematic study of its impact on \emph{minimum cloud latency} for users, which is becoming an increasingly important factor in determining user experience. This latency is determined by the location of the nearest cloud edge, and that is under control of the major cloud providers. Some important questions regarding cloud edge locations and their impact on the digital divide include: Are the cloud providers considering the digital divide in choosing these edge locations? Would the digital divide increase or decrease with these cloud locations? What can be done to potentially reduce the digital divide?  


In this paper, we attempt to answer these questions -- we use globally distributed probes to measure the improvements possible with the cloud edge, and combine measures of economic well-being with edge datacenter locations to analyze the extent of the digital divide with respect to cloud latency. To the best of our knowledge, we are the first to study how the deployment of cloud edge affects global access to high quality Internet services. As part of our study, we make three primary contributions:

First, we perform a measurement study to quantify the reduction in latency due to the cloud edge. We perform the study on the widely used measurement platform RIPE Atlas~\cite{ripe_atlas}. For targeting cloud edge locations, we initially choose the Local Zone product of AWS, and then broaden our results by using cloud routing optimization services to \emph{predict} edge latency for three major cloud providers. Our results in \S\ref{sec:cloud_edge_latency} show up to 19\% latency reduction at the 80th percentile in the US, and \emph{greater than 80\% of probes in North America had cloud edge access under the 20ms threshold (which is required for head-tracking)}. The latency reduction in under-provisioned continents was even higher; however, there is still a large gap between these continents and more developed regions even when using the cloud edge.

Second, in \S\ref{sec:digital_divide} we demonstrate how cloud edge widens the digital divide in datacenter distances through metrics widely used in development economics. Using global population rasters~\cite{world_pop} and existing or announced AWS datacenters, we observe that the \emph{inequality} (which we define as the ratio between those furthest and those closest to datacenters) doubles for many continents with the introduction of the cloud edge. Oceania, in particular, has the top 90\% of users over 140x further from a datacenter than the bottom 10\%. Using measures of datacenter distance combined with economic indicators including census data~\cite{updated_demographics} and night time lights~\cite{viirs_ntl}, we show this inequality is also unfair. This analysis quantifies the extent to which cities hosting cloud edge locations tend to have higher amounts of wealth. We offer suggestions for how to select more optimal locations which minimize unfairness while reaching the most users.

Lastly, we zoom in on the latency and digital divide when considering ISPs using low Earth orbit (LEO) satellites (e.g., Starlink). We hypothesize that, just like it has improved global access to the Internet, it may be a promising technology to improve inequality in access to the nearest cloud as well. This is based on intuition that delay of the satellite hop dominates the end-to-end delay and is fairly homogeneous across the globe. Our case study in \S\ref{sec:leo_satellites} focuses on answering the following question: Can LEO Internet reduce the digital divide in cloud datacenter access? Through measurements with RIPE Atlas probes using Starlink connectivity, we first show their feasibility in providing low latency cloud access suitable for tasks requiring the edge. Next, we take into account distances on the satellite hop, and find that the differences in cloud distance between the top and bottom 10\% of users drops to lower than 10x. By widening the area reachable by a single local zone, unfairness stemming from the selection of cities also greatly declines.

These results demonstrate that the digital divide, previously observed to disadvantage communities without Internet access, now separates communities by an increasingly important metric - minimum cloud latency. As reliance on the cloud increases for everyday tasks, and applications such as AR~\cite{tooba-ar-chi, tooba-ar-cscw} or remote work~\cite{netflix_local_zones} require lower latency and improved cloud access, users on one side of this digital divide may be left behind. We believe our work is a first step in highlighting this issue, and provides guidance on promising technologies and deployment paths that could help in reducing this divide.

\section{Background}

In this section we provide background on the relevant technologies that are the focus of this paper. This includes cloud datacenters and their private WANs as well as satellite internet service providers.

\subsection{Cloud Networks}\label{sec:background_cloud_networks}

Multiple cloud providers operate large private WANs to connect their regions directly to user ISPs while bypassing the public Internet. The extent of private WANs allow cloud providers to reach over 76\% of the internet without using Tier-1 or 2 ISPs~\cite{flat_internet}. Due to this "flattening" an increasing portion of the network path between users and cloud resources is falling under cloud provider's ownership. For example, a California based user accessing a server in Amazon’s Virginia datacenter can expect ~50ms round trip time (RTT), most of which is in Amazon’s private WAN that spans the country.

These private WANs are known to not have a large effect on minimum cloud latency within continents that have a well provisioned public internet infrastructure~\cite{cloudy}. However, they do offer significant quality advantages particularly over long distances~\cite{measuringwidearea, jqos, rewan}. The increased reliability from these networks is now a commercial product for many cloud providers. For example, Google offers two tiers of network service with the paid tier using a private network as much as possible for ingress and egress~\cite{private_wan_performance}. Additionally, AWS offers Global Accelerator~\cite{aws_global_accelerator} to route ingress through the closest AWS PoP~\cite{private_wan_performance}. To reduce latency beyond the limits imposed by the physical distances of routing even through these private networks, cloud providers have begun pushing datacenters towards the network edge. These new cloud edge datacenters are smaller than traditional regions, and offer a subset of services with different pricing models. Local Zones are the AWS edge compute product and it has seen significant deployment in recent years~\cite{local_zone_locations}. Each Local Zone is connected to a parent region over the private WAN.

 \begin{figure}
    \centering
    \includegraphics[width=1\columnwidth]{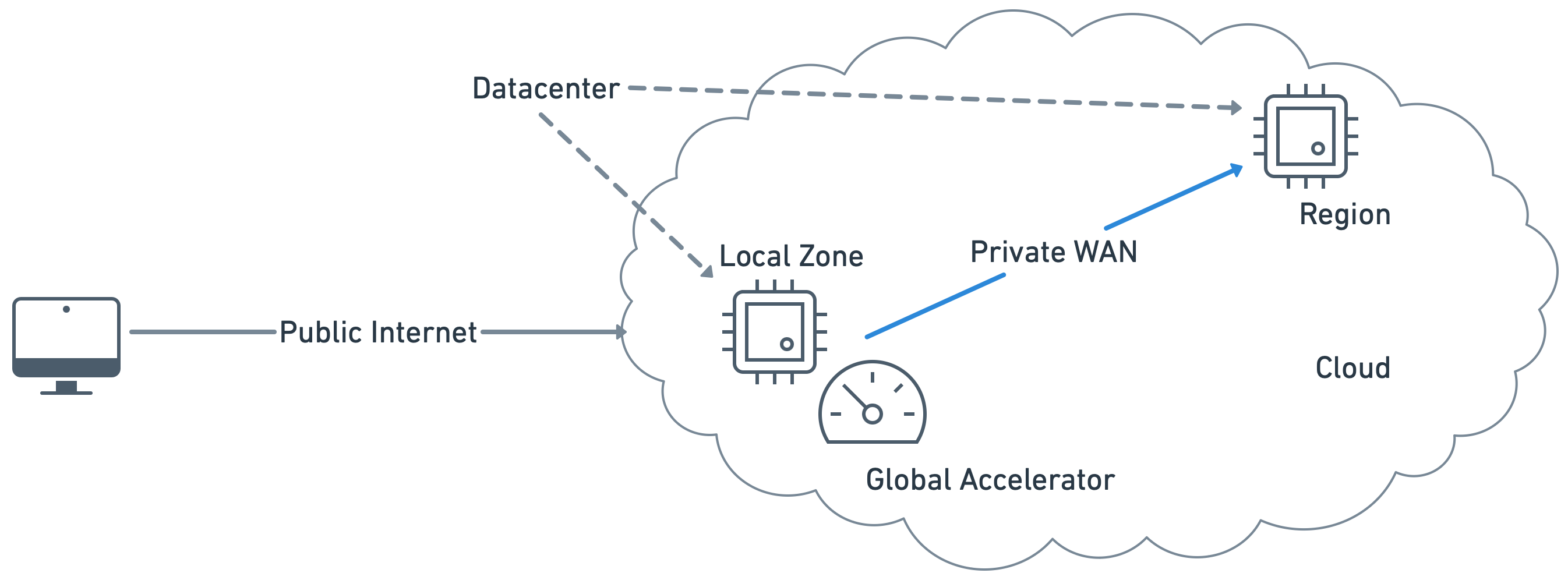}
    \caption{Structure of Amazon's services and terms used in this paper. Local Zones are datacenters at the edge of the cloud, and are connected to traditional regions over the private WAN. Global Accelerator, the routing optimization service, allows traffic on the public internet to maximize usage of the private WAN to reach datacenter regions.}
    \label{fig:experiment_structure}
\end{figure}

Fig~\ref{fig:experiment_structure} shows the structure of Amazon's services we are describing. Within the cloud is a private network that connects two kinds of datacenters: regions and local zones. When users connect to the cloud they first use the public Internet. If they are accessing the nearest cloud edge the request reaches its destination without traversing the private WAN. If they are accessing a further datacenter, such as a cloud region, the private WAN can be used for improved performance. The Global Accelerator service ensures network traffic stays in the private WAN as much as possible even when accessing a datacenter a long distance away. This is done using an anycast IP address advertised from multiple edge locations. TCP connections are then split between the public Internet that connects the user and the Global Accelerator service, and the private WAN connecting Global Accelerator to the datacenter~\cite{aws_global_accelerator_tcp_split}.

\subsection{LEO Satellites}

Satellite internet is a promising technology to bring Internet access to people in regions with no ground-based ICT infrastructure. However, geosynchronous satellites incur large RTT overheads due to orbiting at over 35,000 km from Earth~\cite{hypatia}, requiring hundreds of milliseconds of latency even at the speed of light. Internet from Low Earth Orbit (LEO) satellites enables large reductions in latency compared to these geosynchronous satellites. This is because LEO satellites such as Starlink are orbiting less than 600 km away from Earth~\cite{browser_starlink}. To provide enough ground coverage, LEO satellites must operate in large constellations with thousands of satellites~\cite{browser_starlink}. Once deployed, a LEO satellite based ISP can create a path from a customer's satellite dish, through low Earth orbit, to a ground station that is connected to the Internet and cloud datacenters with much lower latency than geosynchronous networks.

\section{Cloud Edge Latency}\label{sec:cloud_edge_latency}

In this section, we present results on a measurement study conducted with the widely used RIPE Atlas platform to answer two question about cloud edge latency. First, what is the latency reduction possible from using the already supported cloud edge datacenters in the United States (\S\ref{sec:us_latency})? Second, how much faster do we expect cloud latency to be when three of the major cloud providers launch their edge compute platforms globally (\S\ref{sec:per_continent_baseline})? We find nearly a 2x improvement in probes that can currently reach Amazon's cloud in under 20ms. Our technique to measure global cloud edge latency predicts a 14\% to 70\% improvement per continent when compared to a baseline (\S\ref{sec:baseline}). In \S\ref{sec:private_wan_extent} we find the prediction technique to be accurate for at least 95\% of probes in each continent.

\subsection{U.S. Cloud Edge Latency}\label{sec:us_latency}

\begin{wrapfigure}{r}{0.5\textwidth}
  \begin{center}
    \includegraphics[width=0.48\textwidth]{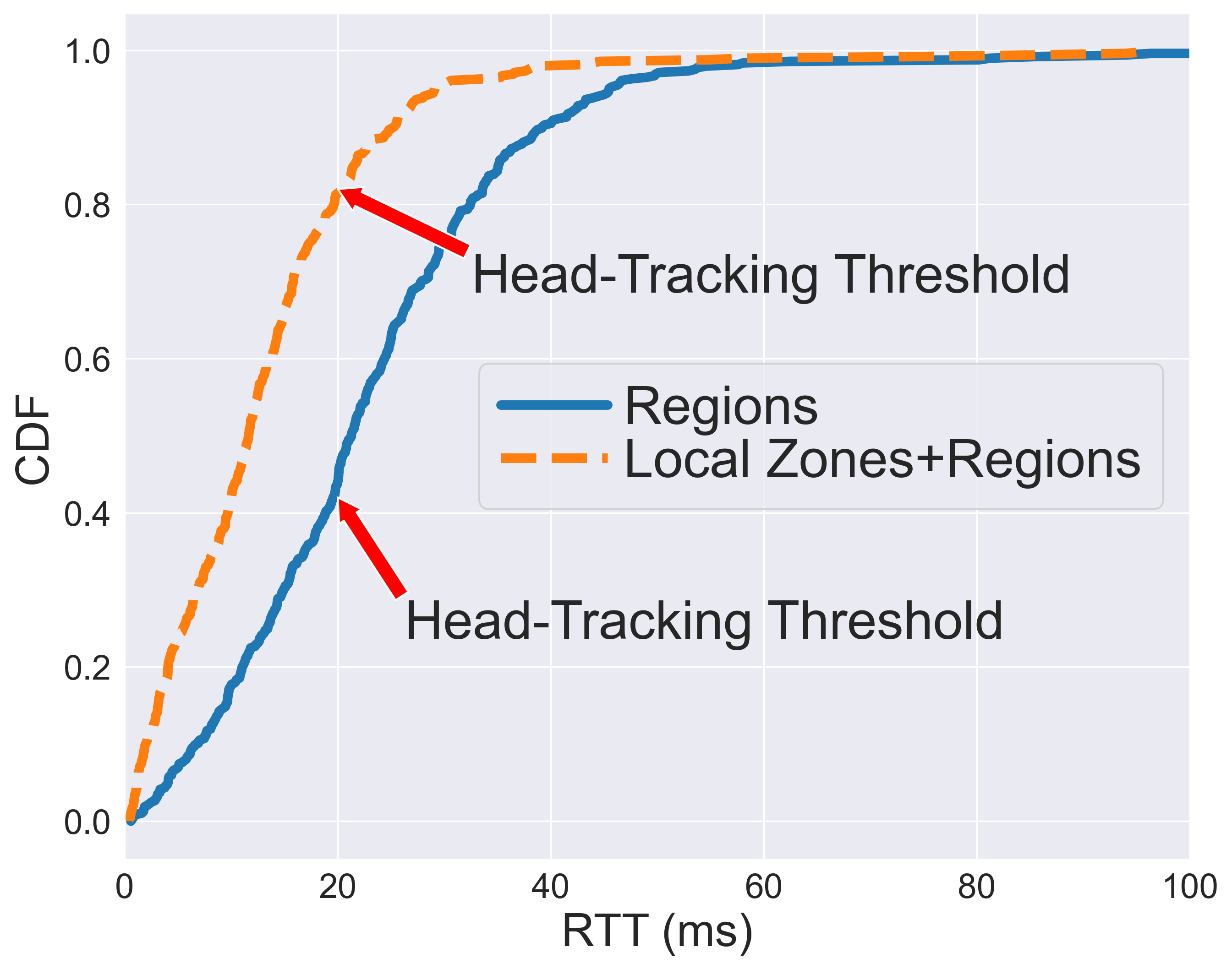}
  \end{center}
  \caption{Latency improvements with Local Zones vs. Regions}
  \label{fig:local_zone_speedup}
\end{wrapfigure}

Only AWS had made their cloud edge service publicly available at the time of our experiments, and their Local Zones were only launched within the U.S. To measure expected improvements for current applications that can use cloud edge, we compare the single cloud provider case of AWS regions only vs. all AWS datacenters.

We measured region latency with virtual machines provided by CloudHarmony~\cite{cloud_harmony} and hosted by AWS. Local Zone endpoints were AWS EC2 t3-medium instances launched in all 16 US zones. Fig~\ref{fig:local_zone_speedup} shows the latency improvement measured with traceroutes to each Local Zone and all 4 regions in the US from 500+ RIPE Atlas probes in the country. In this experiment, and all subsequent results, we first excluded RIPE Atlas probes installed in datacenters by filtering the user-provided tag $datacenter$. While this does not guarantee exclusion of probes hosted in privileged locations, it helps keep the results more representative of end user latency. For each probe we recorded the minimum time to reach a datacenter, and the minimum time to reach a region. Nearly 2x as many probes can reach cloud compute in under the 20 ms threshold when using Local Zones. This result demonstrates the large potential of cloud edge compute to reduce latency when applications can be distributed across multiple locations.

When each of the three major cloud providers are considered, minimum achievable latency is even further reduced, but requires a new measurement method because launching instances at the edge was not yet available for Google Cloud and Azure even in the US. Before presenting those results in \S\ref{sec:per_continent_baseline} we validate a baseline to compare against.

\subsection{Global Baseline}\label{sec:baseline}

\begin{wrapfigure}{r}{0.5\textwidth}
  \begin{center}
    \includegraphics[width=0.48\textwidth]{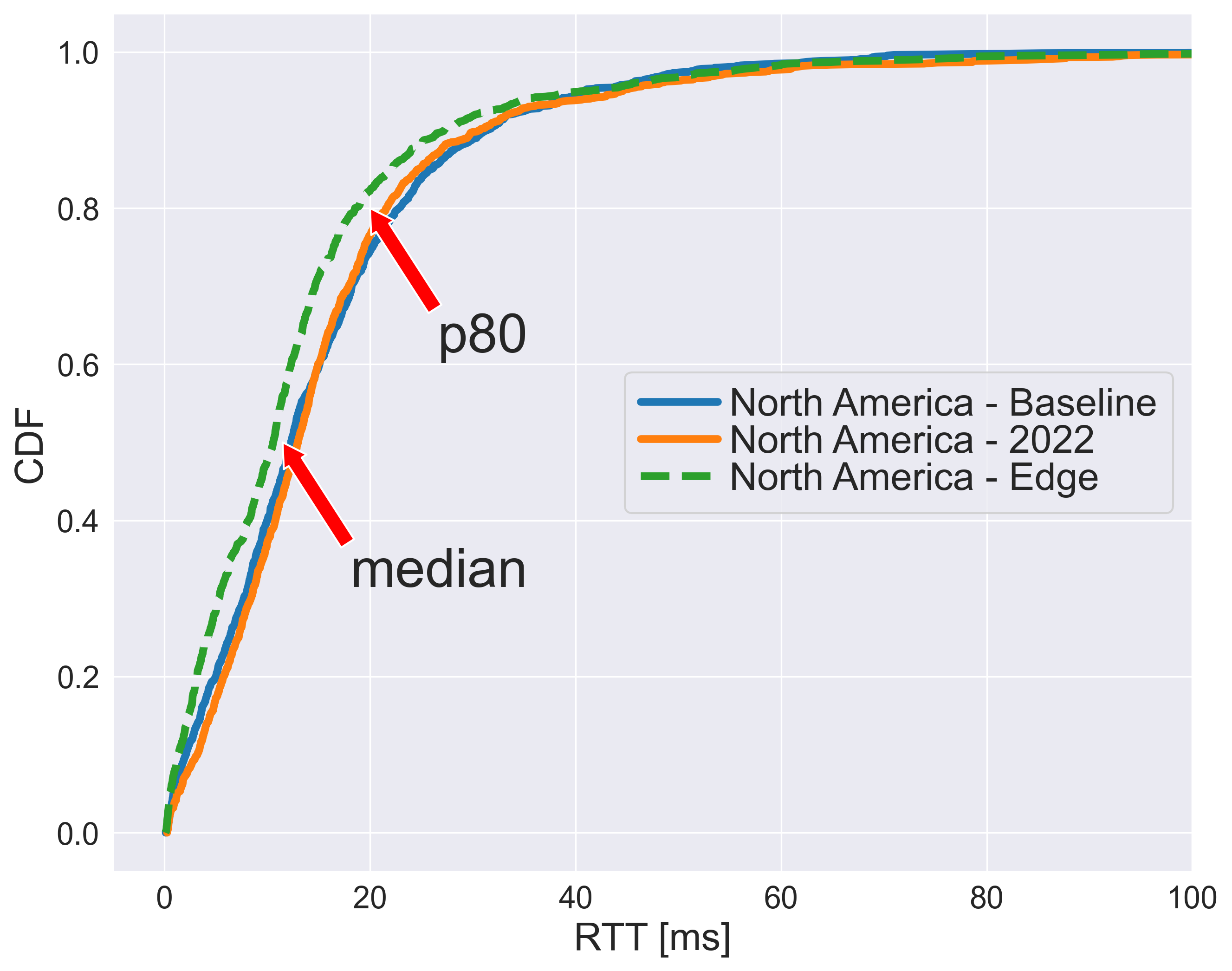}
  \end{center}
  \caption{CDF of round trip time from RIPE atlas probes in North America to 3 destinations: the 2019 baseline, Vultr and Azure regions in 2022, and the cloud edge in 2022. There are not significant changes in the baseline and current cloud regions, but the cloud edge does provide speedups of 19\% p80 and 15\% p50.}
  \label{fig:baselinevalidation}
\end{wrapfigure}

To determine the speedup possible by expansion of the cloud edge, we compare to a previous measurement study of cloud regions performed on RIPE Atlas by Corneo et al in 2019~\cite{surrounded}. Factors other than datacenter location, such as access network latency or WAN improvements, could have changed since the previous study was performed. Therefore, we start by validating latency from our RIPE Atlas probes in North America to the same cloud regions used by the previous study~\cite{surrounded} has remained unchanged since 2019. We performed 3 ICMP pings every 15 minutes for 45 minutes to 20 datacenters representative of those selected for the 2019 measurements. The plot in Fig~\ref{fig:baselinevalidation} shows the minimum latency for each probe as a CDF. The results demonstrate that the current latency measurements closely match those of our baseline. This validates other factors not present at the time of the baseline collection are not influencing latency, and therefore we can use that data to compare with our edge measurements and understand the benefits of the cloud edge.

We predict latency for each of AWS, Google Cloud, and Microsoft Azure by launching endpoints in their routing optimization services known as Global Accelerator~\cite{aws_global_accelerator}, Cloud CDN~\cite{google_cloud_cdn}, and Azure Front Door~\cite{azure_front_door}, respectively. Rather than targeting a specific host from our measurement origin, these endpoints have an anycast IP address that can be reached by the probes. These services offer a higher tier of networking reliability by avoiding the public internet and routing to the private WAN as quickly as possible (\S\ref{sec:background_cloud_networks}). This represents a best case scenario for cloud edge deployments. As more of the cloud edge datacenters are launched, they more closely match the locations that host services such as Global Accelerator.

In contrast to the baseline vs. validation measurements, using the cloud edge does create a significant speedup. The results in Fig~\ref{fig:baselinevalidation} are a 19\% speedup at the 80th percentile and 15\% at the median from using the cloud edge.

\subsection{Predicted Latency Per Continent}\label{sec:per_continent_baseline}

\begin{figure}[!t]
    \centering
    \subfloat[width=0.5\columnwidth][Speedup in NA, EU, OC]{\includegraphics[width=0.495\columnwidth]{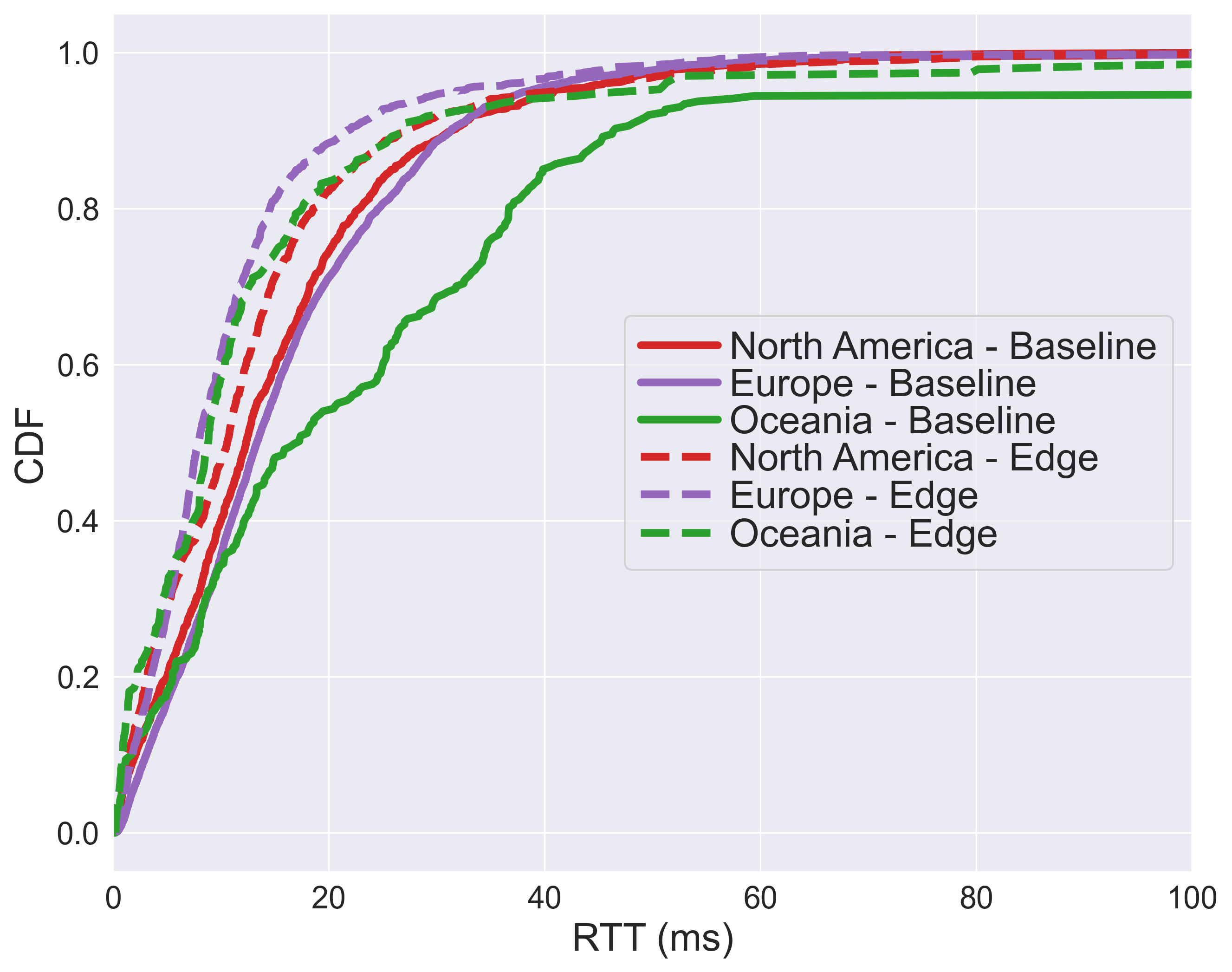}\label{fig:na_eu_oc_speeds}}
    \subfloat[width=0.5\columnwidth][Speedup in AF, SA, AS]{\includegraphics[width=0.495\columnwidth]{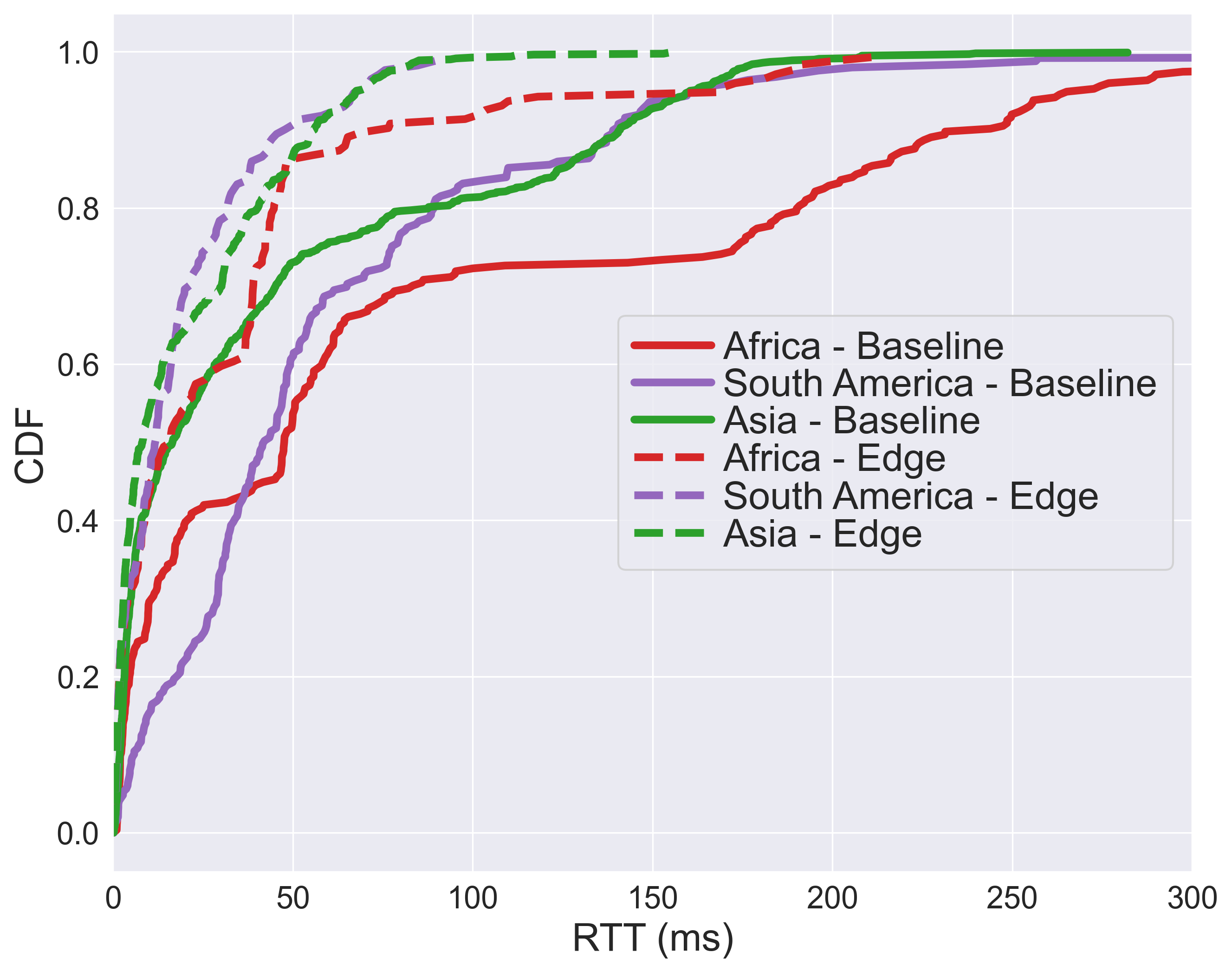}\label{fig:af_sa_as_speeds}}
    \caption{Cloud edge vs. cloud regions for each continent}
    \label{fig:continent_speedup}
\end{figure}

Next, we extend the methodology developed above to find the minimum latency to cloud edge in every continent. Results are again aggregated to the minimum RTT per probe, and compared to our baseline in Fig~\ref{fig:continent_speedup}. Every continent has significantly lower latency to the cloud edge than datacenter regions, and a few trends stand out.

First, all three continents in Fig~\ref{fig:na_eu_oc_speeds} can reach the 20ms threshold for greater than 80\% of probes. These continents - North America, Europe, and Oceania, have sufficient cloud edge infrastructure to support new applications that require the cloud edge for many users. However, as we'll see in \S\ref{sec:world_inequality}, the concentration of RIPE Atlas probes does not always match population, and there can still be high proportions of the population without a nearby cloud edge.

Second, Oceania is particularly notable for the dramatic improvement. New Zealand did not have a datacenter in the baseline; therefore many probes in New Zealand had much higher latency. We speculate this is due to traffic traversing long undersea cables to reach the closest datacenter in Australia. The cloud edge does include datacenters within the island, and sub 20ms latency is achievable in New Zealand.

Lastly, we see in Fig~\ref{fig:af_sa_as_speeds} that the other three continents - Asia, South America, and Africa have even higher improvements, but also a large gap between the fastest and slowest probes. Each of these continents have a greater baseline latency, and Fig~\ref{fig:af_sa_as_speeds} has a larger x-axis range to reflect this. Using the cloud edge makes the median more comparable to the other three continents. In fact, it is consistently under the 20ms threshold. However, the 80th percentile is still significantly slower and cannot support emerging edge applications. Africa's CDF stands out as a bi-modal shape, with about 50\% reaching the cloud edge in 20ms and another 40\% taking >40ms. This indicates a still under-provisioned region that has no nearby cloud edge but high population. We'll return to this idea in \S\ref{sec:inequality} when demonstrating the unequal access between the top and bottom percentiles. Detailed results for each continent's median and p80 are shown in Table~\ref{fig:table}.

\begin{table}[]
\begin{tabular}{l|l|l|l|l|l|l|}
\cline{2-7}
                         & p80 - Baseline & p80 - Edge & Speedup & p50 - Baseline & p50 - Edge & Speedup \\ \hline
\multicolumn{1}{|l|}{NA} & 22.77          & 18.53      & 18.62\% & 12.33          & 10.56      & 14.36\% \\ \hline
\multicolumn{1}{|l|}{EU} & 24.6           & 14.52      & 40.98\% & 13.33          & 7.89       & 40.81\% \\ \hline
\multicolumn{1}{|l|}{OC} & 36.67          & 17.3       & 52.82\% & 16.77          & 8.8        & 47.53\% \\ \hline
\multicolumn{1}{|l|}{AS} & 87.5           & 40.07      & 54.21\% & 16.5           & 8.11       & 50.85\% \\ \hline
\multicolumn{1}{|l|}{SA} & 88.72          & 31.72      & 64.25\% & 41.59          & 11.57      & 72.18\% \\ \hline
\multicolumn{1}{|l|}{AF} & 189.95         & 44.32      & 76.67\% & 47.54          & 15.06      & 68.32\% \\ \hline
\end{tabular}
\caption{Summary of speedups from using cloud edge on each continent}
\label{fig:table}
\end{table}

\subsection{Private WAN Extent}\label{sec:private_wan_extent}

Lastly, we look for exceptions to our assumption that Amazon's Global Accelerator~\cite{aws_global_accelerator} can be used to measure the cloud edge. While all Global Accelerator servers are within Amazon's network and are meant to minimize latency, there still could be other factors preventing the global accelerator endpoint from being the closest AWS infrastructure to a probe.  For example, cloud PoPs may exist in an area to extend the private WAN but not offer the global accelerator service yet due to an incomplete deployment. In this case, a traceroute between probes and Global Accelerator would contain extra hops that would not be necessary if compute was offered at the edge of the cloud private WAN. Cases like these are a limitation in our what-if analysis of the cloud edge, and cause an under-estimation of the possible speedups.

\begin{wrapfigure}{r}{0.5\textwidth}
  \begin{center}
    \includegraphics[width=0.48\textwidth]{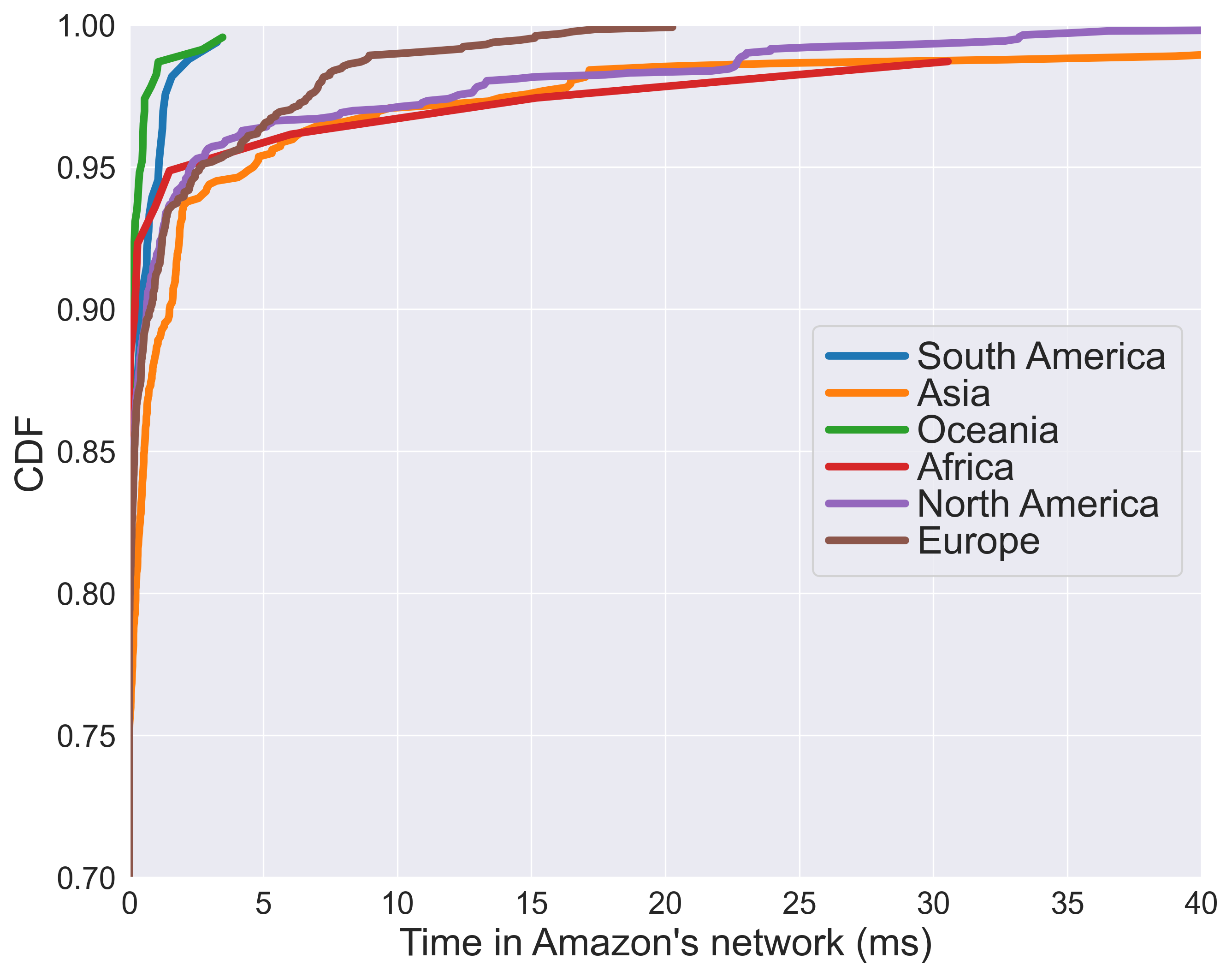}
  \end{center}
  \caption{CDF in difference between first Amazon hop and global accelerator endpoint. Y-axis shows last 30\% of CDF. Some regions have long tails, indicating areas where our method does not capture the lowest possible latency of a completed cloud edge deployment.}
  \label{fig:amazon_time}
\end{wrapfigure}

To quantify this effect, we identify the first hop in a traceroute that is in an ASN operated by AWS. The time in Amazon's network is the difference in minimum time to the first AWS owned hop and the Global Accelerator endpoint. Traceroutes are processed using the python library $cymruwhois$~\cite{cymruwhois} to lookup Autonomous System Number (ASN) and $IPWhois$~\cite{ipwhois} to lookup the owning entity. Fig~\ref{fig:amazon_time} plots CDFs of the time spent in Amazon's network for each continent. Only 5\% of each continent show a significant time in the private WAN, and some continents have none at all. Plotting the affected probes geographically (not shown for brevity) shows a few clusters where the AWS private network exists but there are no Global Accelerator endpoints. One such location was the Philippines. While there is a planned Local Zone in Manila, latency from nearby probes were consistent with the distance to the nearest region, Hong Kong. When the Manila Local Zone is available, we expect these probes to have lower latency than predicted by our what-if analysis. However, we don't expect this to influence the main results as it is isolated to a few locations with low density of probes.

\setlength\intextsep{0pt}

\section{Digital Divide}\label{sec:digital_divide}

The existence of a ``digital divide'' has been well documented, particularly in terms of availability of internet access on a per-country level~\cite{pew_digital_divide}. With the deployment of cloud edge, we already measured a divide between continents with fast access such as North America, Europe, and Oceania and those with significantly higher round trip times. In this section, we analyze the prevalence of a digital divide within continents from two perspectives. 

First, by treating edge datacenters as general purpose compute that can be used from anywhere, we show the existence of a widening gap between those who are closest to a datacenter and those who are furthest. This is the typical way cloud regions are used, but not cloud edge. Local Zones are marketed as a service to be used by people in the same city as the datacenter. However, we believe this is not a fundamental technical limitation, but a common current use case due to the higher costs of Local Zones. By viewing cloud edge as fundamentally the same as other datacenters, just closer to some users, we show in \S\ref{sec:inequality} the inequality in datacenter access is exacerbated by the expansion of edge locations.

Second, we combine measures of datacenter distance with economic indicators to show this inequality is also unfair. This analysis uses the fact that cloud edge locations, while available to anyone, are specifically targeted to their host cities, which we show to typically be cities with higher amount of wealth. In \S\ref{sec:unfairness} we explore the relationship between which cities are selected and the economic well being of those cities to demonstrate an unfair trend of higher income cities having access to the cloud edge.

\subsection{Inequality in the Edge}\label{sec:inequality}

Inequality in cloud access occurs when there are significant differences in minimum latency to datacenters. The most important factor in determining minimum latency is typically distance~\cite{cloudy}, and reducing distance is the primary advantage of deploying new edge datacenters. Therefore, we quantify inequality in the distribution of distance to the nearest datacenter. We focus on the percentile ratio, specifically $p90/p10$, which captures how much access is improved by the cloud edge for the fastest 10\% ($p10$) compared to the slowest 10\% ($p90$). While there are many ways to measure inequalities, this statistic is commonly used in economics to study income~\cite{census_inequality_increased, ibr_income_inequality} and can be intuitively extended to datacenter distances.

\begin{figure}[!t]
    \centering
    \subfloat[Initial (21)]{\includegraphics[width=0.30\textwidth]{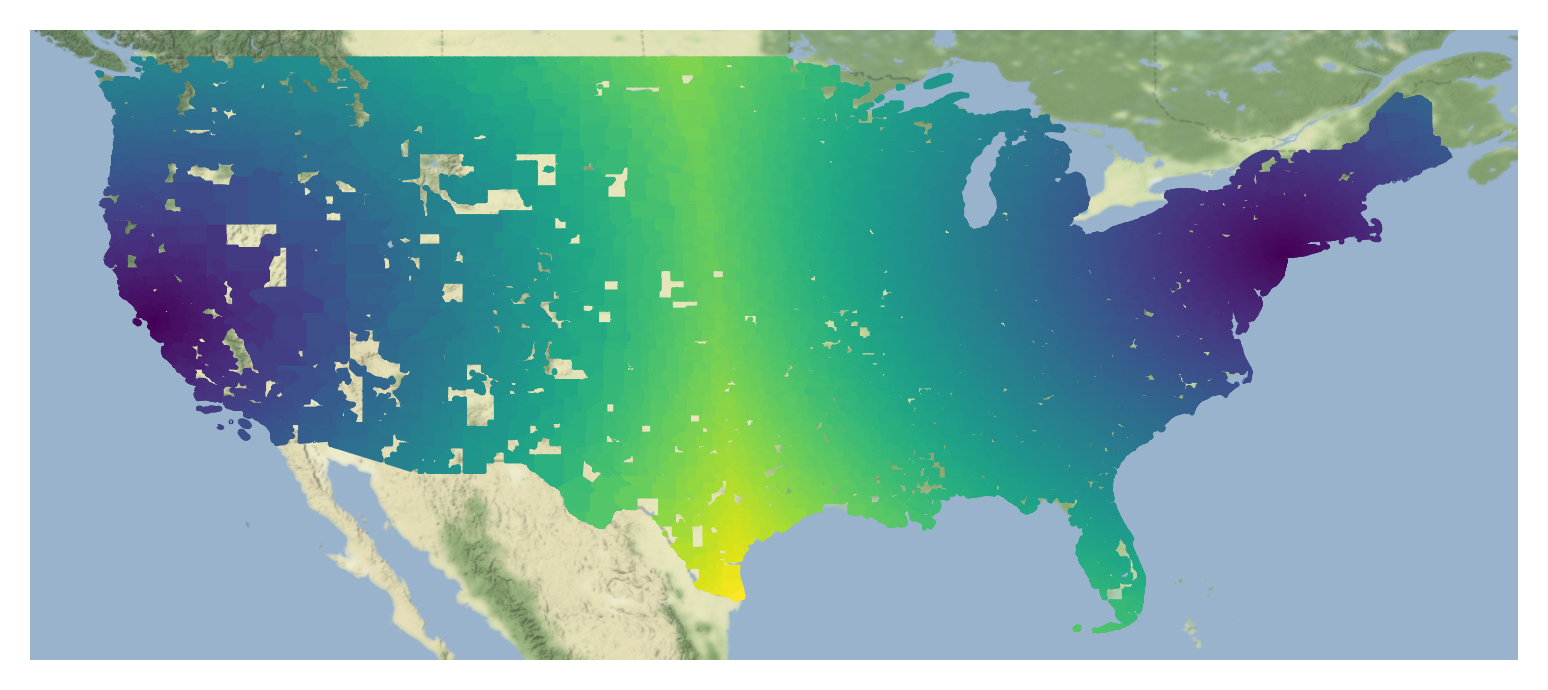}\label{fig:initial_percentile}}
    \subfloat[Higher Inequality Location (46)]{\includegraphics[width=0.30\textwidth]{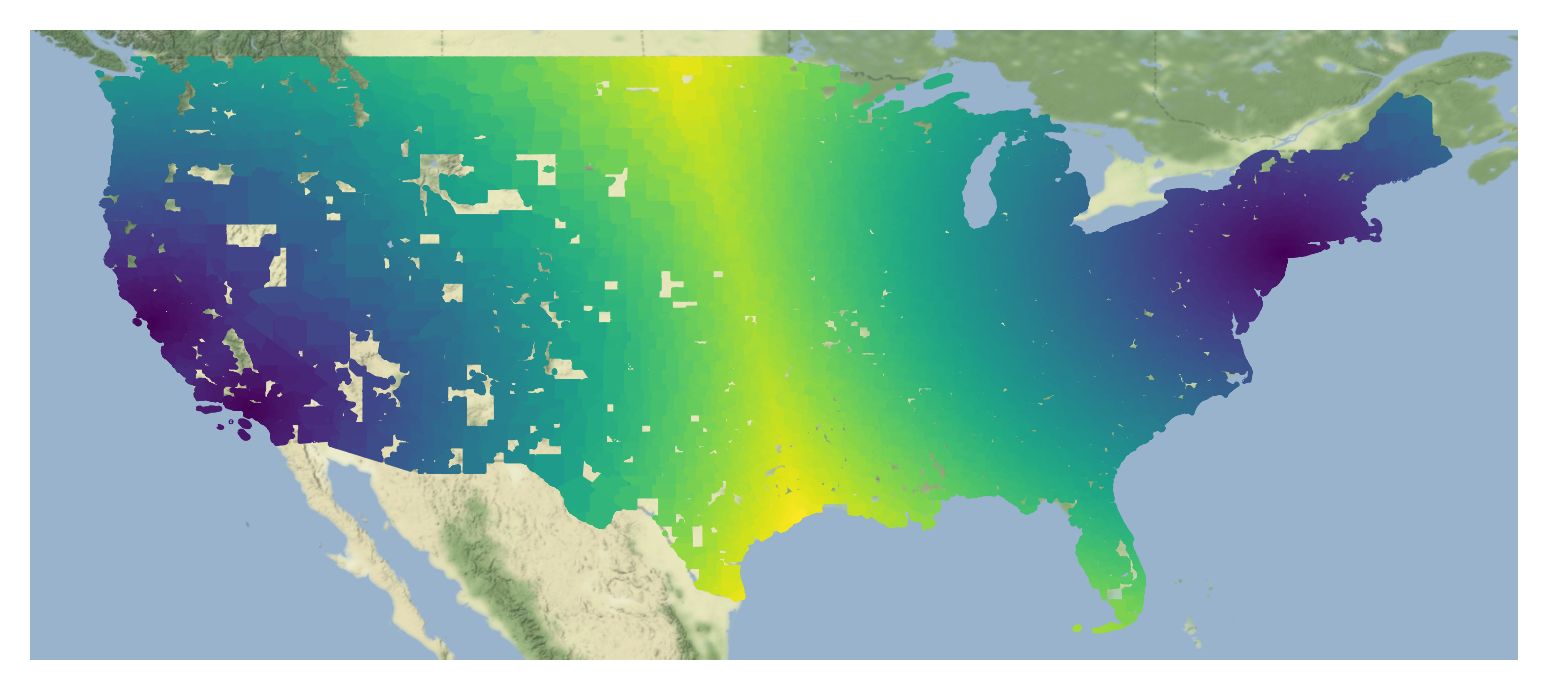}\label{fig:unfair_percentile}}
    \subfloat[Lower Inequality Location (16)]{\includegraphics[width=0.30\textwidth]{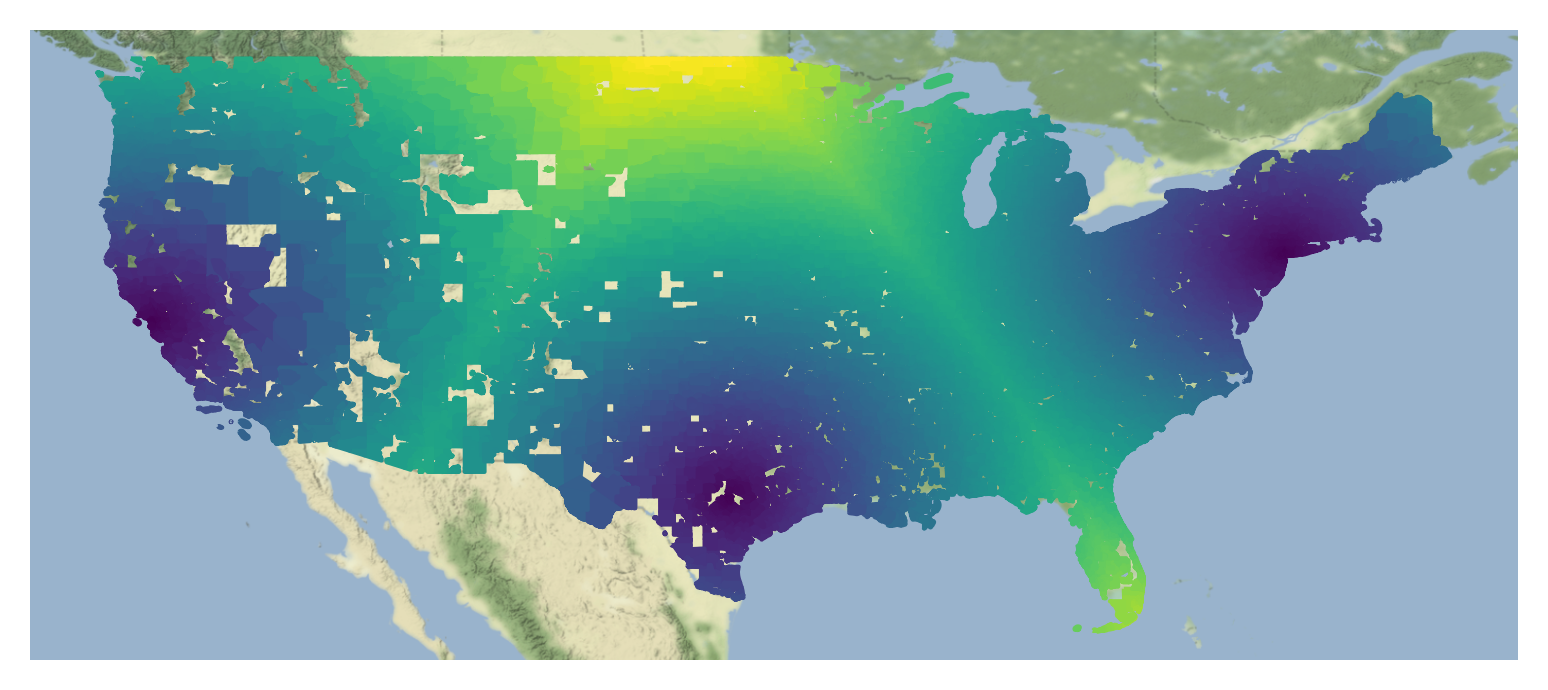}\label{fig:fair_percentile}}
    \caption{Toy example of increasing and decreasing inequality with datacenter locations and the associated percentile ratios. \ref{fig:initial_percentile} starts with a datacenter in San Francisco and New York City and a p90/p10 of 21. \ref{fig:unfair_percentile} adds a location in Los Angeles which increases the p90/p10 to 46 because this location was already close to a datacenter. \ref{fig:fair_percentile} adds a location in Austin, TX which decreases the p90/p10 to 16.}
    \label{fig:latency_gap_toy}
\end{figure}

Fig~\ref{fig:latency_gap_toy} presents a toy example illustrating our method of calculating inequality, using datacenter locations in the US. Fig~\ref{fig:initial_percentile} shows distances to the initial datacenter locations on the west coast (San Francisco) and east coast (New York City). Fig~\ref{fig:unfair_percentile} is the same map with one new datacenter added in Los Angeles. This location increases inequality because it reduces distances in locations that were already relatively close to a datacenter ($p10$), but not for people who were far ($p90$). Fig~\ref{fig:fair_percentile} decreases inequality by placing a new datacenter in Austin - which decreases distances for the furthest population.

\subsubsection{US Inequality}\label{sec:us_inequality}
As more Amazon edge locations have been added in the US, inequality has been rising. Starting from datacenter regions the $p90/p10$ inequality is 9.1. In this case users in the 90th percentile are less than 10x further from a datacenter than users in the 10th percentile. With currently launched Local Zones included as datacenter regions, that number rises to 23.1. In the US, there is only one edge location that does not already have a Local Zone (Jacksonville, Florida). This city is relatively close to others with datacenters such as Miami and Atlanta, which explains the slight increase in inequality to 27.8 that we expect if a datacenter launches in Jacksonville.

\begin{figure}
    \centering
    \subfloat[CDF of distances]{\includegraphics[width=0.45\textwidth]{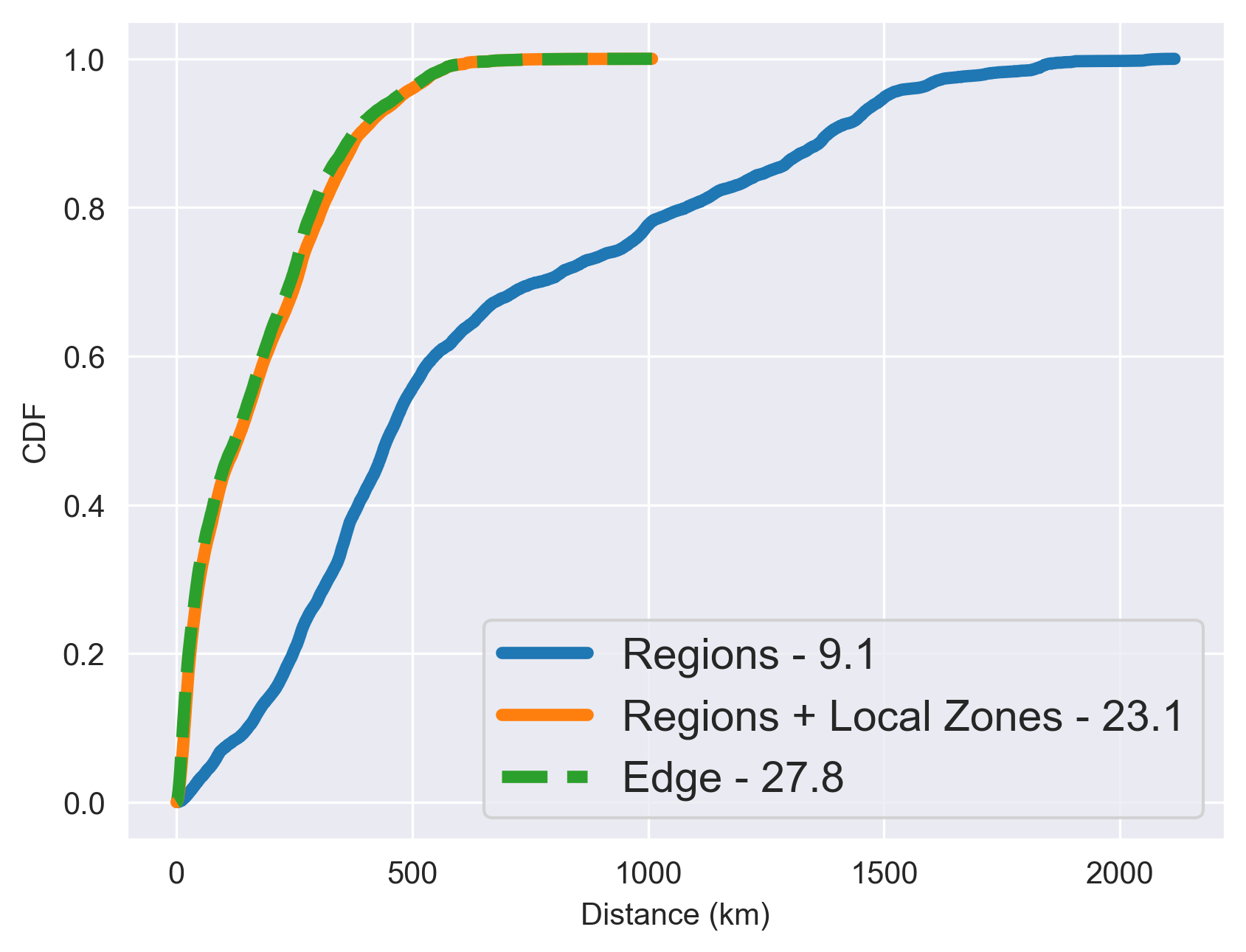}\label{fig:us_percentiles_cdf}}
    \subfloat[p90/p10 as new edge datacenters are launched]{\includegraphics[width=0.45\textwidth]{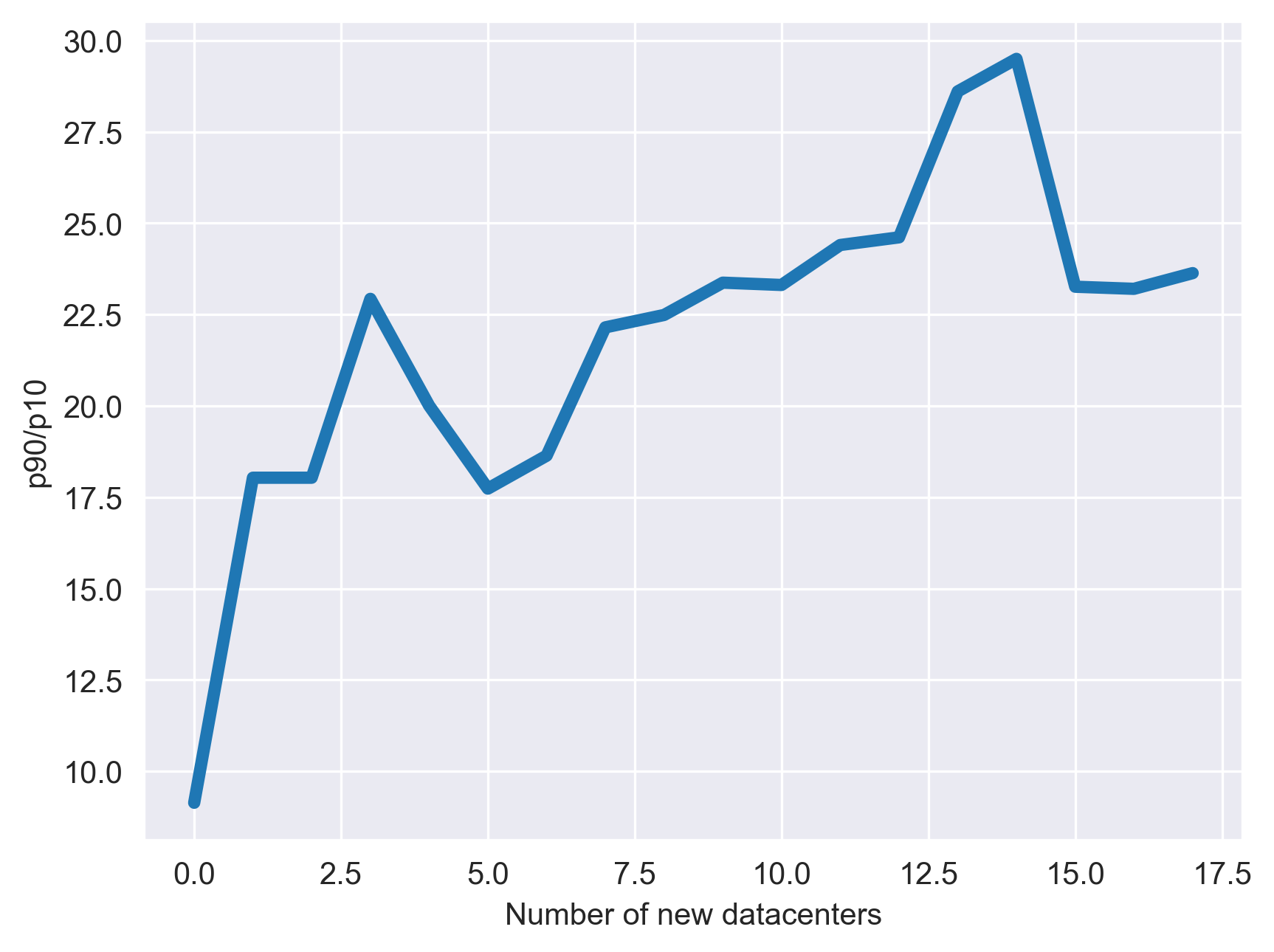}\label{fig:us_percentiles_time}}
    \caption{Measures of inequality in U.S. datacenter placements. \ref{fig:us_percentiles_cdf} compares the CDF of datacenter distance from populations of census tracts for regions, regions and Local Zones, and all edge locations. Inequality as measured by the p90/p10 increases when the cloud edge datacenters are available. \ref{fig:us_percentiles_time} plots the p90/p10 as each new Local Zone was launched. In general, new launches increase this measure of inequality.}
    \label{fig:us_percentiles}
\end{figure}

New datacenters at the cloud edge decrease the distance for everyone, but they often launch in areas that are already close to existing datacenters. This causes the overall inequality to increase. The CDFs in Fig~\ref{fig:us_percentiles_cdf} demonstrate this trend. When edge locations are included, the closest users get even closer, but the users furthest from datacenters are left on a longer tail. Fig~\ref{fig:us_percentiles_time} demonstrates the change in inequality with each new local zone launch -- most of these cases increase the $p90/p10$ with the exception of two large drops when Miami and Atlanta are added. These cities were in locations that previously had no nearby datacenter. The trend points to decreasing latency for an increasing proportion of people but some users are left behind and won't be able to access the same low-latency applications as the rest of the country. This further creates a digital divide.

\textbf{Details on the datasets.} We computed these results using population counts  from the American Community Survey (ACS) and downloaded from the  Economic and Social Research Institute (ESRI) Updated Demographics~\cite{updated_demographics} using ArcGIS Online~\cite{arcgis_online}. Population is grouped at the census tract level and latitude/longitude for each tract is downloaded from the US census TIGER FTP archive for 2019~\cite{tiger}. These coordinates are used to determine the distance from each tract to datacenters.

\subsubsection{World Inequality}\label{sec:world_inequality}
\begin{figure}
    \centering
    \subfloat[p90/p10 and p80/p20 both increase when compute is expanded to the edge]{\includegraphics[width=0.50\textwidth]{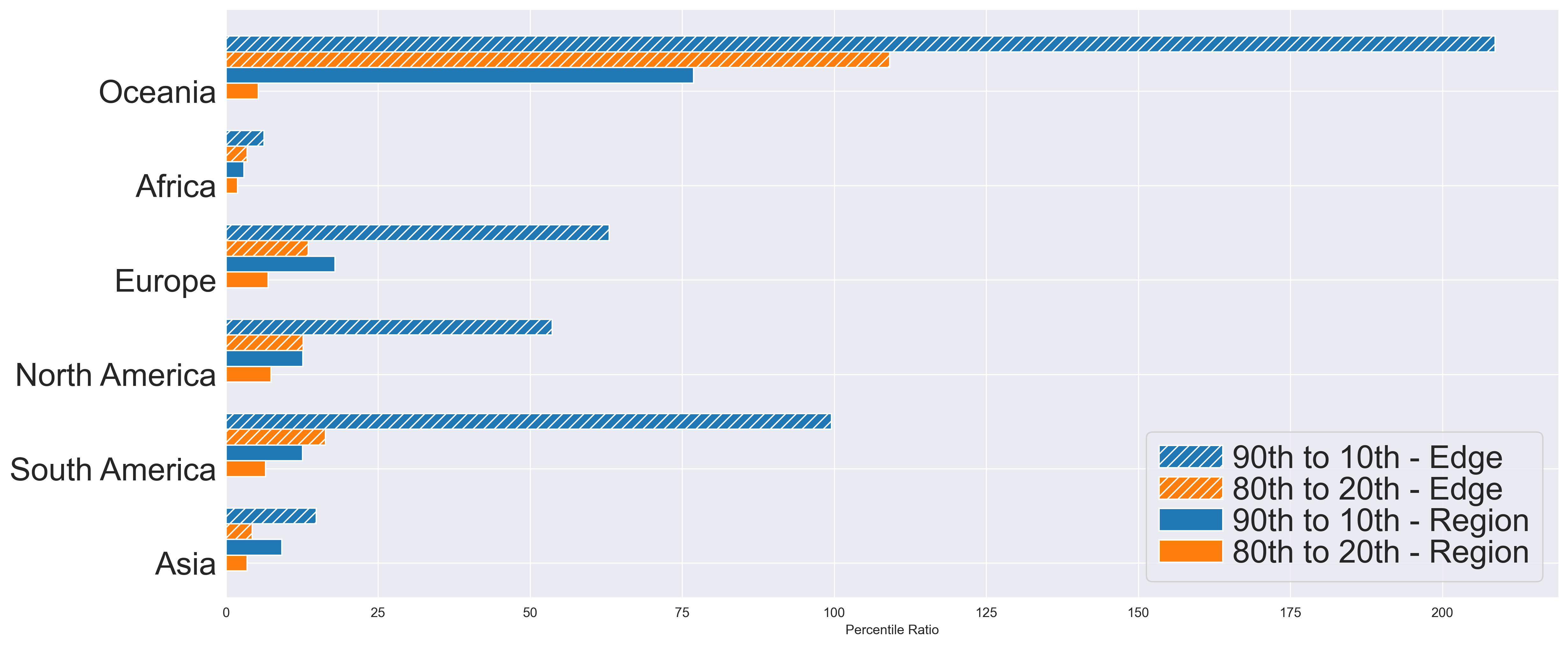}\label{fig:distance_gap}}
    \enspace
    \subfloat[Change in p90/p10 per continent as new datacenters are launched. For all continents the inequality has increased since the first launch of a Local Zone.]{\includegraphics[width=0.45\textwidth]{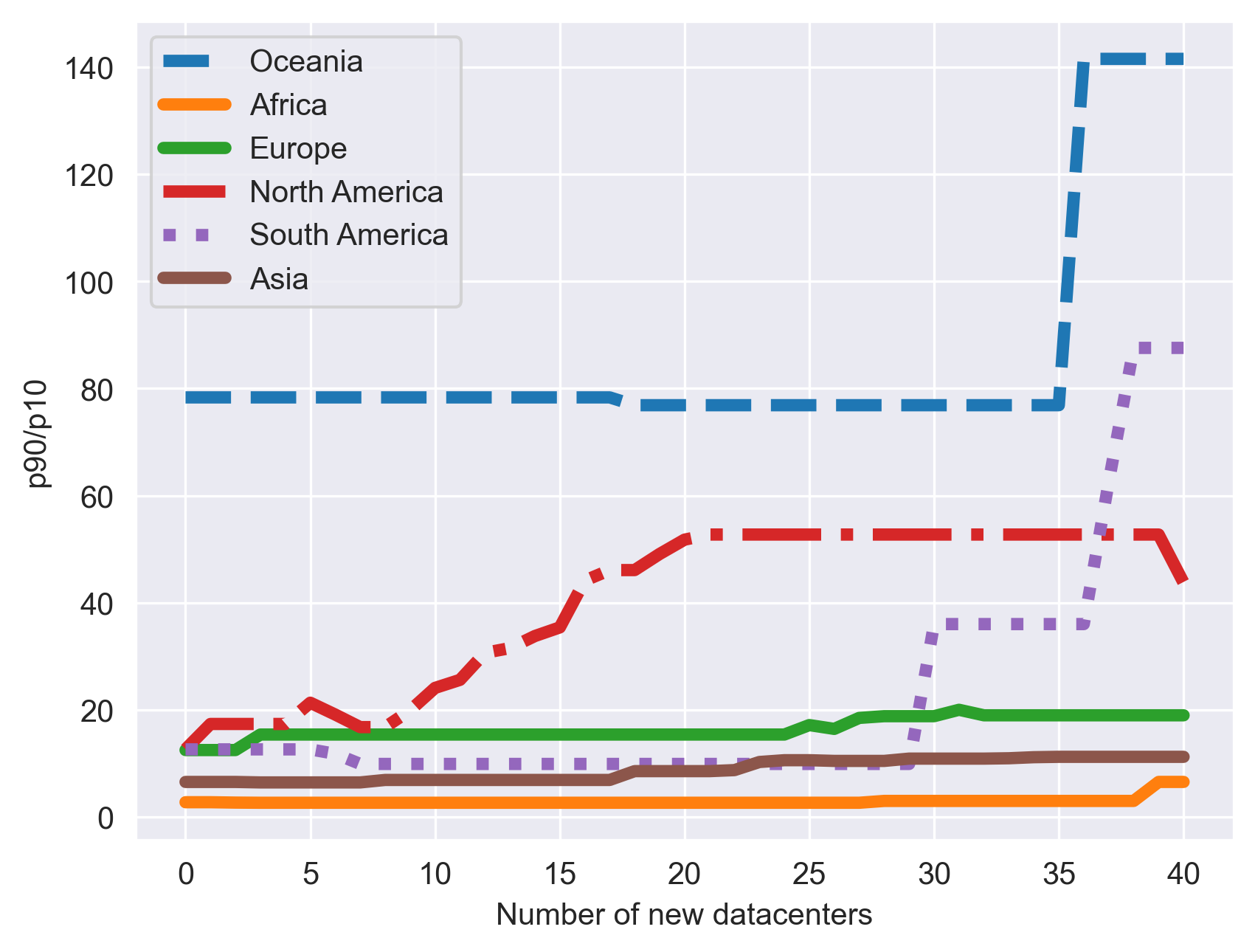}\label{fig:global_percentiles_time}}
    \caption{Inequality measured in percentile ratios for each continent.}
    \label{fig:world_percentiles}
\end{figure}

Globally, the deployment of Local Zones is not yet complete, but we use announced locations and existing edge locations to see how inequality will increase as more are launched.
Fig~\ref{fig:global_percentiles_time} shows the change in p90/p10 as new datacenters are launched. For all continents it has increased since the first Local Zone was introduced. One of the largest jumps was Oceania, which added new locations in Australia and New Zealand. This greatly decreased latency for many in those countries, but did little to help other islands that were already far from the AWS regions. There is also a slight drop in North America with the most recent launch. This was the first location added in Mexico, and helps reduce the p90/p10 because it lowers distance for people who were previously located much further away from datacenters.

While the $p90/p10$ looks at differences in the extremes, we can also examine other ratios, such as $p80/p20$, to see if the increasing trend still applies. Fig~\ref{fig:distance_gap} plots $p90/p10$ as well as $p80/p20$ for each continent when considering all edge locations and regions only. Both the $p90/p10$ and $p80/p20$ increase with the addition of edge locations in every continent.

\textbf{Details on the datasets.} Population for these results are the unconstrained UN-adjusted population counts provided by WorldPop for 2020~\cite{world_pop}. These population values per country are adjusted so the sum matches the UN estimate. Data was downloaded as raster files for each country where the value of each pixel is the population count of an approximately 100m x 100m square. To reduce the data size, we pre-processed the rasters to  5\% of the original size using $gdalwarp$~\cite{gdalwarp} and the $sum$ resampling method to preserve total population counts. Before computing inequality the population was grouped at the Administrative 1 level. This aligns with the groupings used in \S\ref{sec:global_unfairness}. Boundaries of Adaministrative 1 regions were downloaded from from the Database of Global Administrative Areas (GADM)~\cite{gadm}. The goal of GADM is to contain detailed administrative boundaries for all countries, but some are not yet available so these were excluded from our analysis.

\subsection{Unfairness in the Edge}\label{sec:unfairness}

While inequalities measure the difference between the closest and furthest populations, unfairness occurs when the furthest populations have this disadvantage due to factors beyond their control. The digital divide has been shown to unfairly affect communities based on various attributes including race~\cite{whoGetsAccessLA} and poverty~\cite{characterizingCaliforniaInternetQuality}. We use a measure of unfairness commonly used for measuring health, the concentration index, to show the amount of unfairness in datacenter distributions. Measuring in this way requires defining a metric analogous to those used in health, e.g. life expectancy or number of hospitals. We introduce our metric: cloud access indicator in \S\ref{sec:concentration_index} to capture the availability of edge datacenters.

\subsubsection{Concentration Index}\label{sec:concentration_index}

Concentration curves are used to asses bivariate inequities, or unfairness, in a health metric by demonstrating the relationship between the chosen metric and a socio-economic attribute. The curve is defined as the cumulative percentage of the health metric on the y-axis, and the cumulative percentage of population, ranked by the socio-economic attribute, on the x-axis~\cite{analysing_health_equity}. The concentration index (CI) quantifies unfairness and is defined as twice the area between the line of equality (y=x) and the concentration curve~\cite{analysing_health_equity}. This statistic has been used in previous studies to demonstrate unfairness in resource distribution such as access to medial treatment~\cite{health_inequality_hospital_beds} or measurements of population health~\cite{health_inequality_measurements}, but to the best of our knowledge has not previously been applied to datacenter distributions.

Multiple Local Zones are already being offered in the same city for added capacity and resilience~\cite{aws_second_los_angeles}. To capture this in our measure of unfairness, we define cloud access indicator (CAI) for a population (p) and set of datacenter locations ( $L$) as our population health metric:
\begin{equation}
CAI(p) = \sum_{l \in L} reachable(p, l)
\end{equation}
\begin{equation}
reachable(p, l) = \begin{cases}
1 &\text{distance(p, $l$) <= $\sigma$}\\
0 &\text{distance(p, $l$) > $\sigma$}
\end{cases}
\end{equation}

Where the $distance$ function determines the kilometers between a datacenter location and a population. With this metric, each datacenter can contribute to the "health" of a population if it is within a distance of $\sigma$. $\sigma$ represents the maximum distance of a datacenter to be considered helpful to a community. A large value allows more datacenters to factor into the health score for a particular area. Setting it to $\infty$ would make all datacenters equally useful to everyone. By varying this parameter, we can measure the fairness for different cloud applications. Applications requiring low latency will need a lower $\sigma$. We set $\sigma = 70 km$ to measure fairness of applications where users are located in the same city as the datacenter and expect to have very low latency. In \S\ref{sec:leo_satellites} we show how results change when varying $\sigma$.

A completely fair concentration index has a value of 0, and occurs when the entire population has an equal number of datacenters within $\sigma$. The maximum unfairness favoring high income communities has a value of 1, and occurs when only the highest income person has nearby datacenter access. Our dataset groups users into administrative regions so this extreme case would not be possible, but it is approximated by only the highest income region having nearby datacenter access. It is also possible to have maximum unfairness favoring low income communities. This is a concentration index of -1, and occurs when only the lowest income individual has a nearby datacenter. In this case, the concentration curve would be above the line of equality.


\subsubsection{U.S. Unfairness}\label{sec:us_unfairness}

\begin{figure}
    \centering
    \subfloat[Population, log scale]
    {\includegraphics[width=0.30\textwidth]{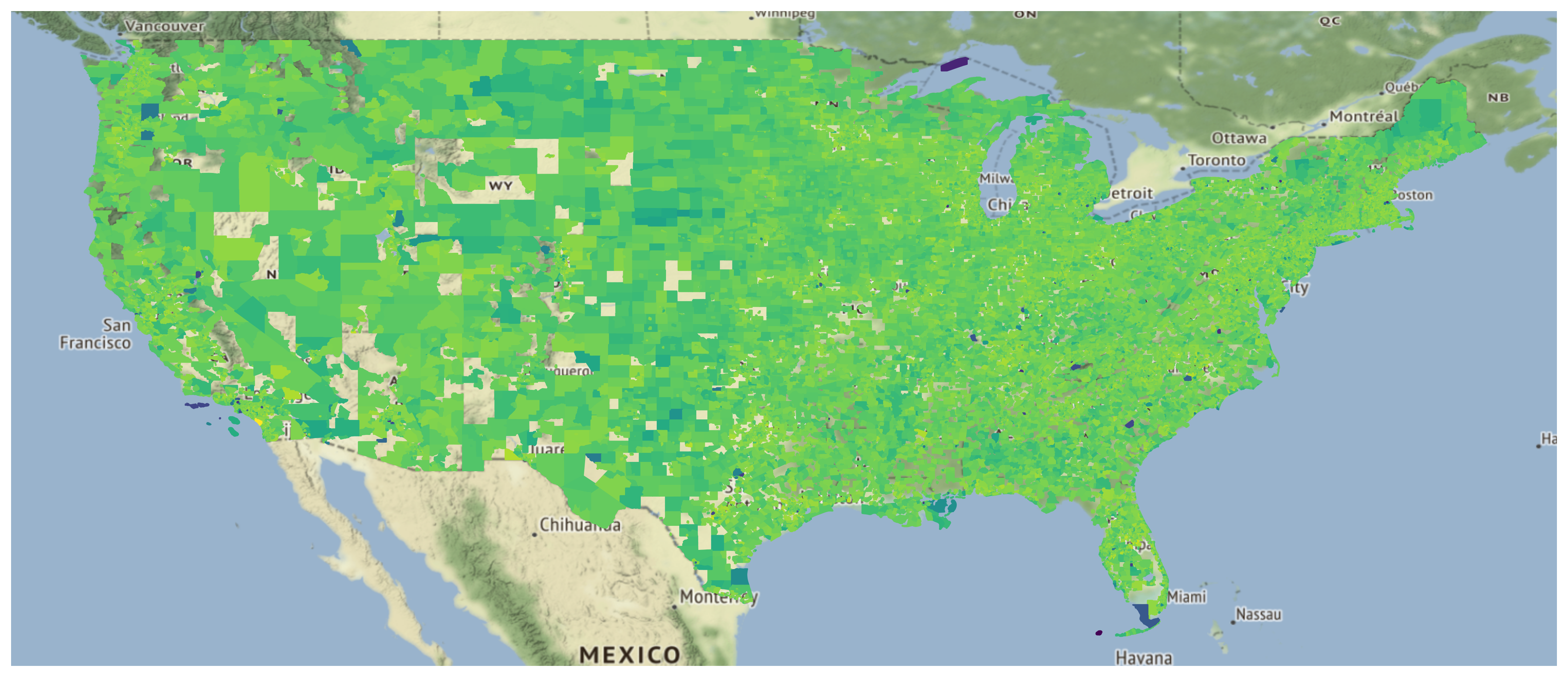}\label{fig:map_pop}}
    \subfloat[Median income, log scale]
    {\includegraphics[width=0.30\textwidth]{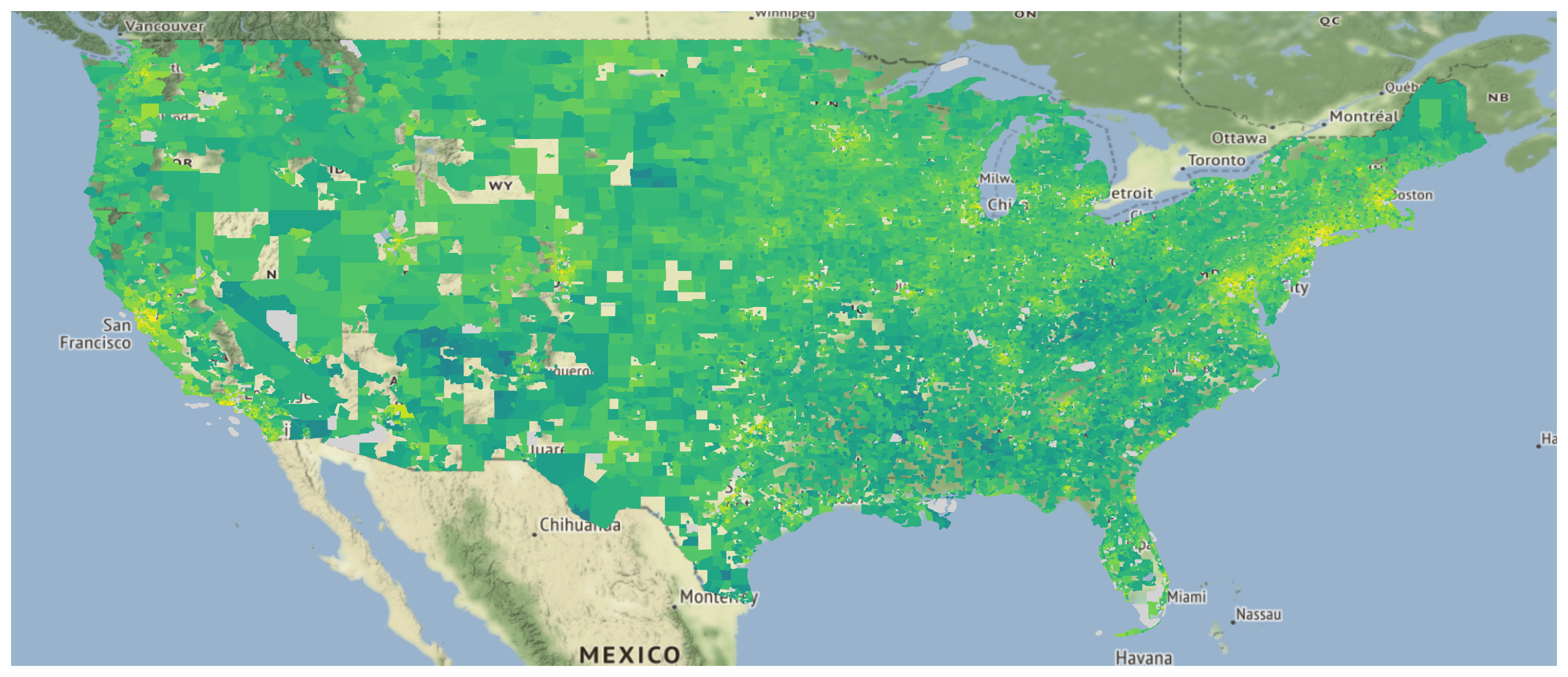}\label{fig:map_income}}
    \subfloat[CAI whem all US region and Local Zones are available]
    {\includegraphics[width=0.30\textwidth]{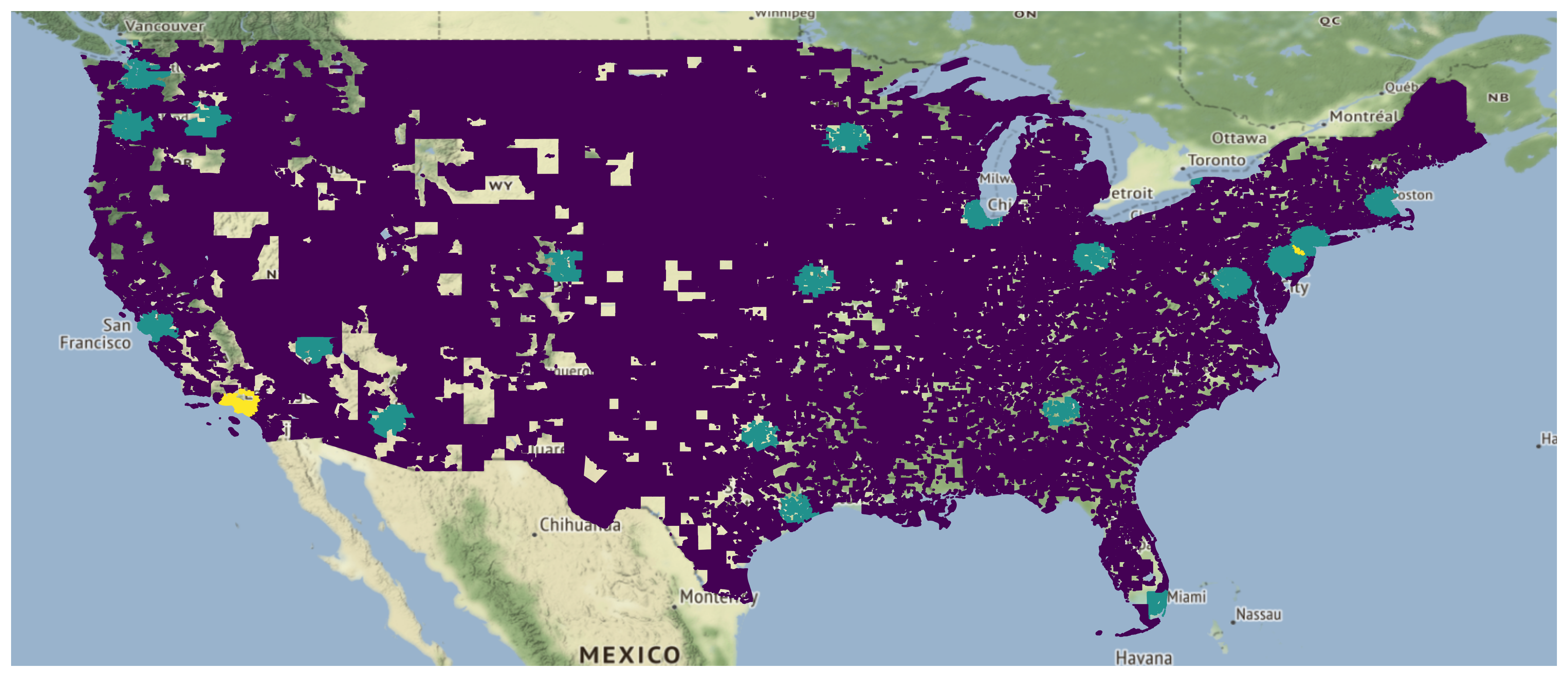}\label{fig:map_cai}}
    \caption{Maps of variables used to calculate concentration curves in the U.S. Each variable is computed for a census tract.}\label{fig:us_maps}
\end{figure}

Within the US, we perform our analysis on population aggregated by census tract. There are 3 variables for each census tract that affect the unfairness: population and median income from the ACS as well as CAI. The spatial distribution of each variable is visualized in Fig~\ref{fig:us_maps}. These maps are generated for datacenter locations in AWS regions, regions and Local Zones, and edge locations.

\begin{figure}
    \centering
    \subfloat[width=0.45\textwidth][Concentration curves of US census tracts for AWS regions and Local Zones.]
    {\includegraphics[width=0.45\textwidth]{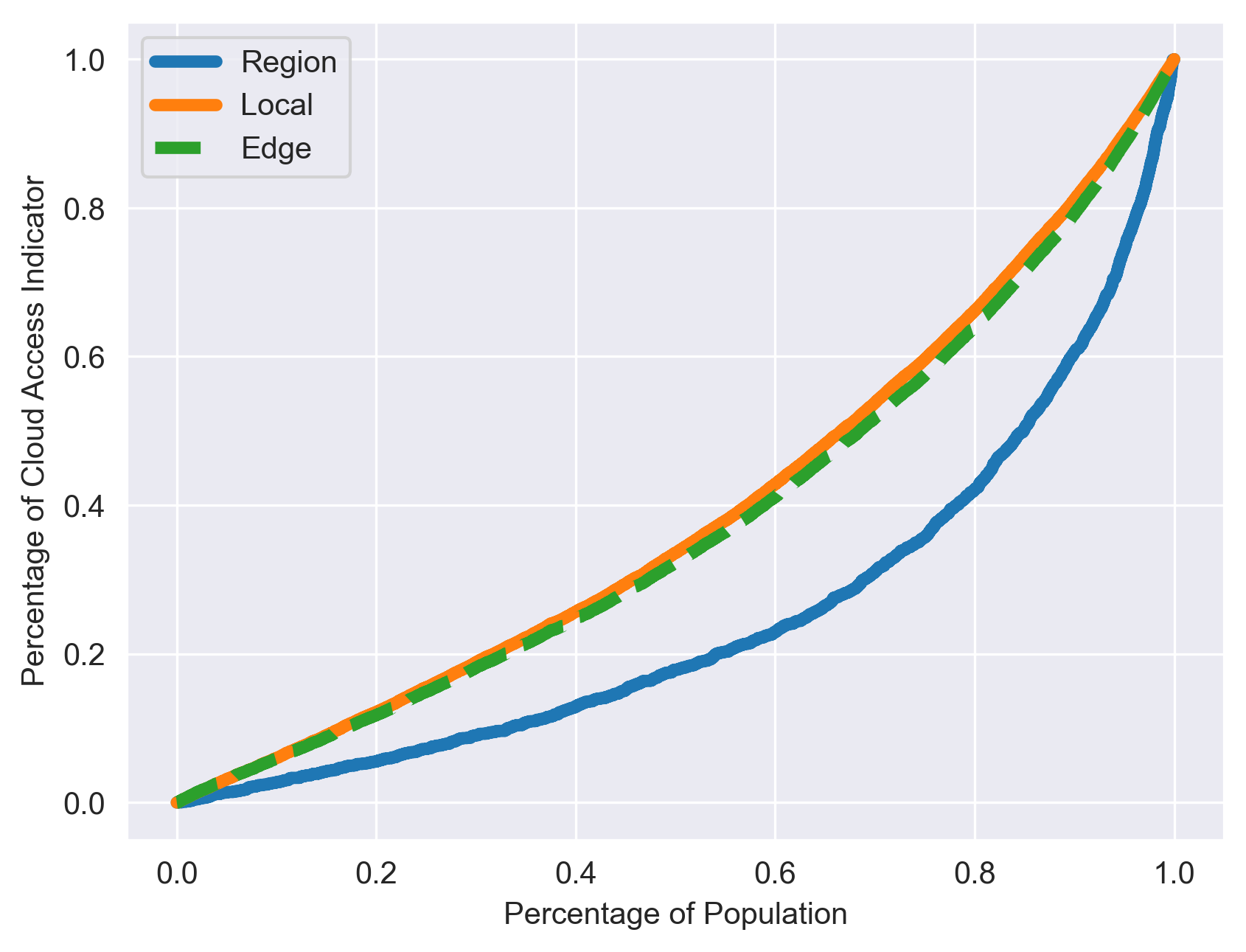}\label{fig:concentration_curves_us}}
    \enspace
    \subfloat[width=0.45\textwidth][Concentration index change as new datacenters are added.]
    {\includegraphics[width=0.45\textwidth]{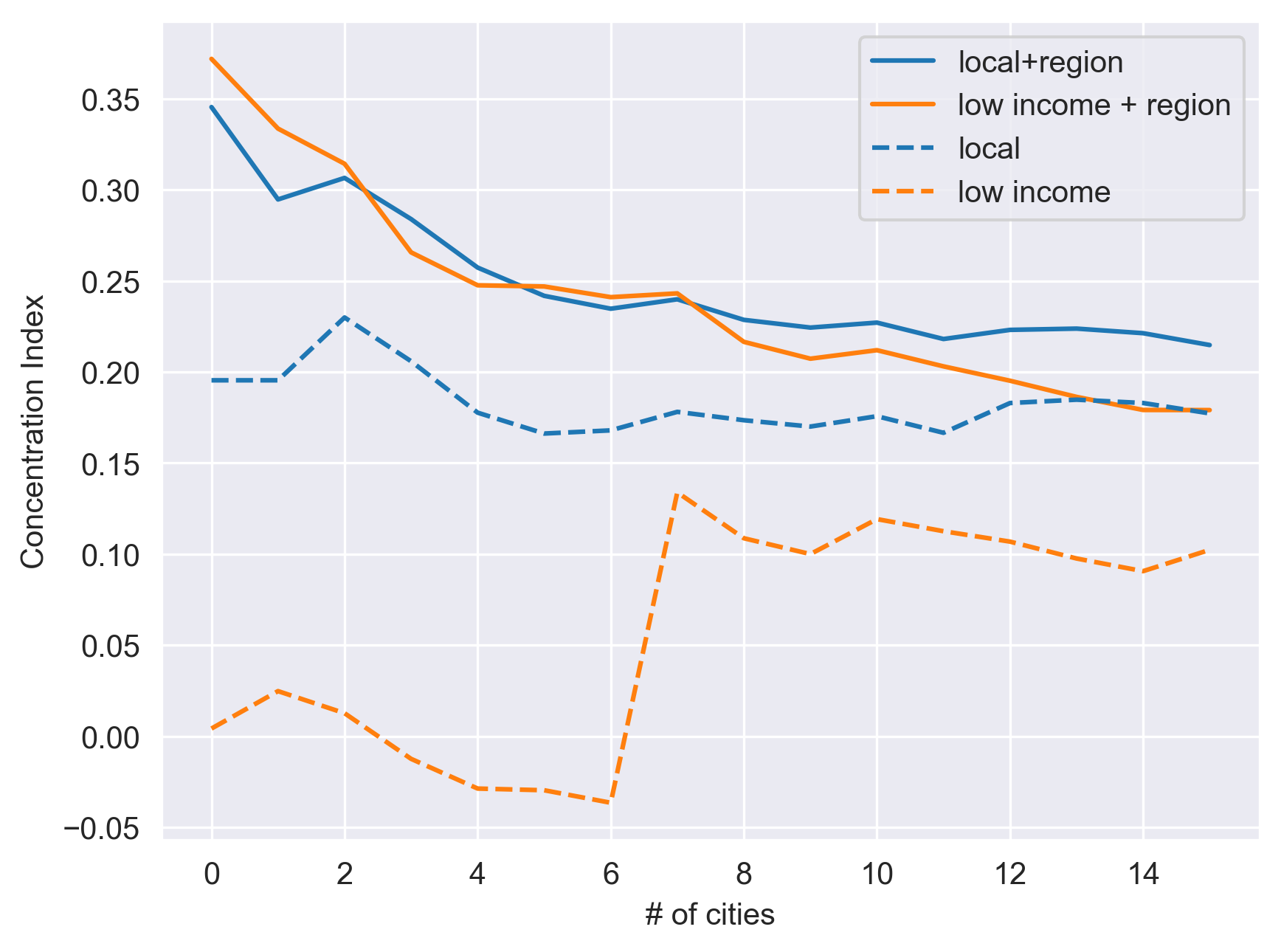}\label{fig:concentration_index_us}}
    \caption{Unfairness in CAI ($\sigma = 70km$) represented by the concentration index at the level of US census tracts.}
\end{figure}

In all three cases, there is a greater availability of datacenter resources in higher income communities, creating unfairness. This is graphically represented in Fig~\ref{fig:concentration_curves_us} as the concentration curve being below the line of equality. The cloud edge does help reduce this unfairness, due to additional cities with local zones containing lower income census blocks. The concentration index for regions is 0.50, for regions + Local Zones it is 0.22, and for the additional city with an edge location it is 0.24.

The deployment of datacenters is already unfair before any locations are added on the network edge. A main contributor to this is the us-east-1 region which is located in silicon valley - containing many of the census tracts with the highest median incomes. In \S\ref{sec:optimal_selection} we elaborate on this with comparisons among major US cities. Local Zones are not particularly fair locations, they are just more fair than this unfair starting point of regions. Fig~\ref{fig:concentration_index_us} demonstrates the change in concentration index as more datacenter locations are added to cities. This plot compares AWS Local Zone deployments (the blue lines), and a hypothetical deployment that adds the lowest income US cities (orange lines). The solid lines include regions at the starting point, and the dashed lines are new locations only. The Local Zone only line shows that each city keeps the CI around 0.2. As more Local Zones are added to the region locations, the CI approaches this value. However, the hypothetical low income city deployment demonstrates that a lower CI can be achieved. Without regions, there is a large gap between the fairness or these two deployment strategies. When the lowest income cities are included in the AWS region locations, the CI still diverges from the Local Zone value, but only after around eight cities are added to overcome the unfairness of the initial regions.

\subsubsection{Global Unfairness}\label{sec:global_unfairness}

\begin{figure}
    \centering
    \subfloat[Mean NTL in South America Administrative 1 regions, log scale]
    {\includegraphics[width=0.30\textwidth]{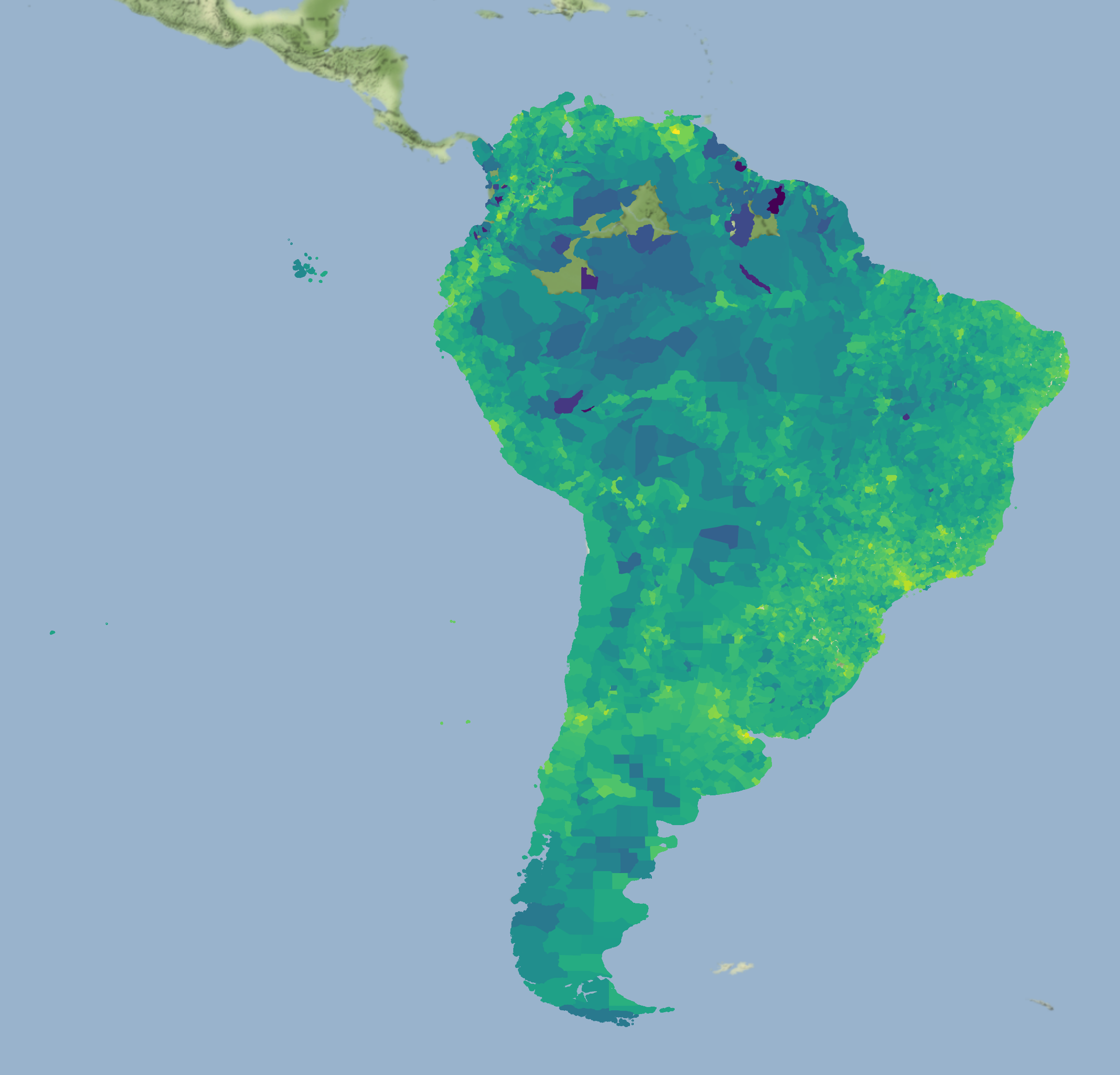}\label{fig:map_lights}}
    \enspace
    \subfloat[Population in South America Administrative 1 regions, log scale]
    {\includegraphics[width=0.30\textwidth]{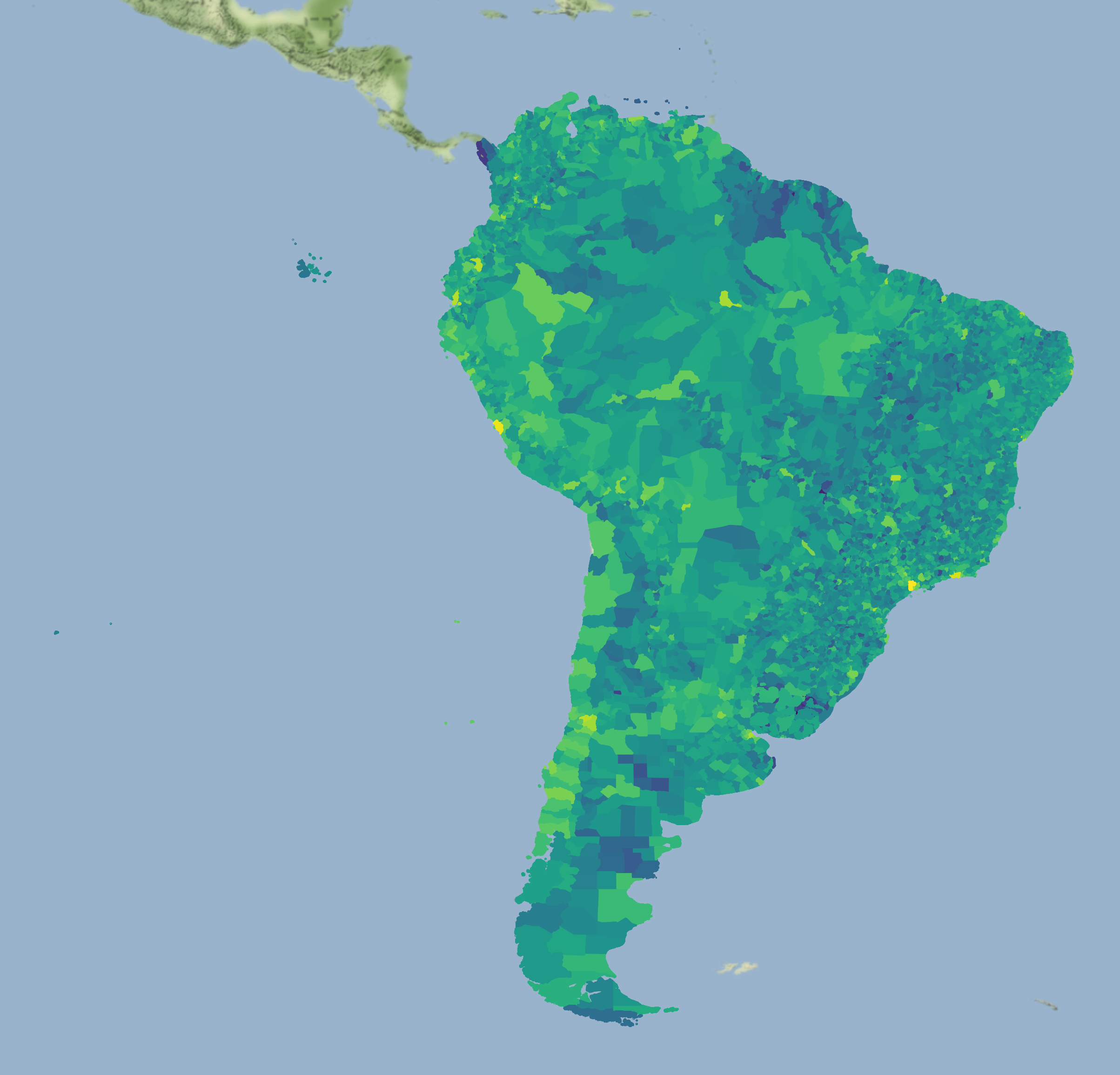}\label{fig:map_pop_sa}}
    \enspace
    \subfloat[Distance to nearest Local Zone in South America]
    {\includegraphics[width=0.30\textwidth]{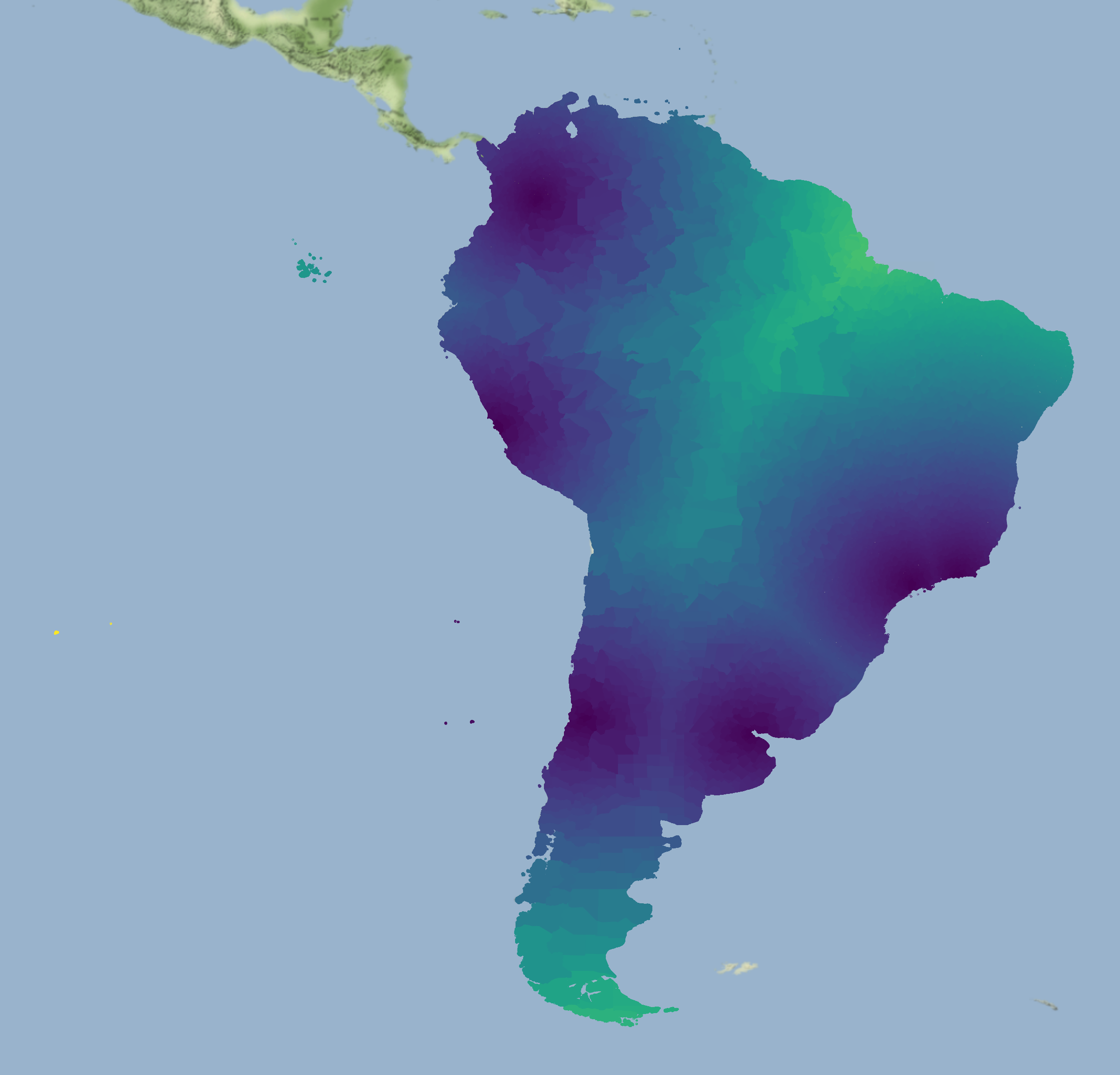}\label{fig:map_distance_sa}}
    \caption{Data for South America}
\end{figure}

Globally there is not a comprehensive source of economic data to sort regions in place of the ACS used in \S\ref{sec:us_unfairness}. However, remotely sensed nighttime lights (NTL) have been shown to be a suitable proxy for economic well-being \cite{night_lights, ntl_economic_development}. This data source is global, on the same scale for every country, and not subject to embellishment for political purposes~\cite{night_light_dictators}. We retrieve NTL rasters from Annual VNL (VIIRS Nighttime Lights) V2.1 \cite{viirs_ntl} for the year 2020. This is a composite of monthly NTL datasets that removes clouds, sunlight, moonlight, and other outliers such as biomass burning. The raster is scaled to match the size of our population raster. We use the mean NTL per administrative 1 unit, based on findings in~\cite{night_lights}, as our proxy for the area's wealth. Mean NTL is calculated with the ``zonal\_stats'' function of the rasterstats python library. Fig~\ref{fig:map_lights} demonstrates the NTL dataset, grouped into administrative regions, for South America.

\begin{figure}
    \centering
    \subfloat[Concentration Curves - Regions]
    {\includegraphics[width=0.32\textwidth]{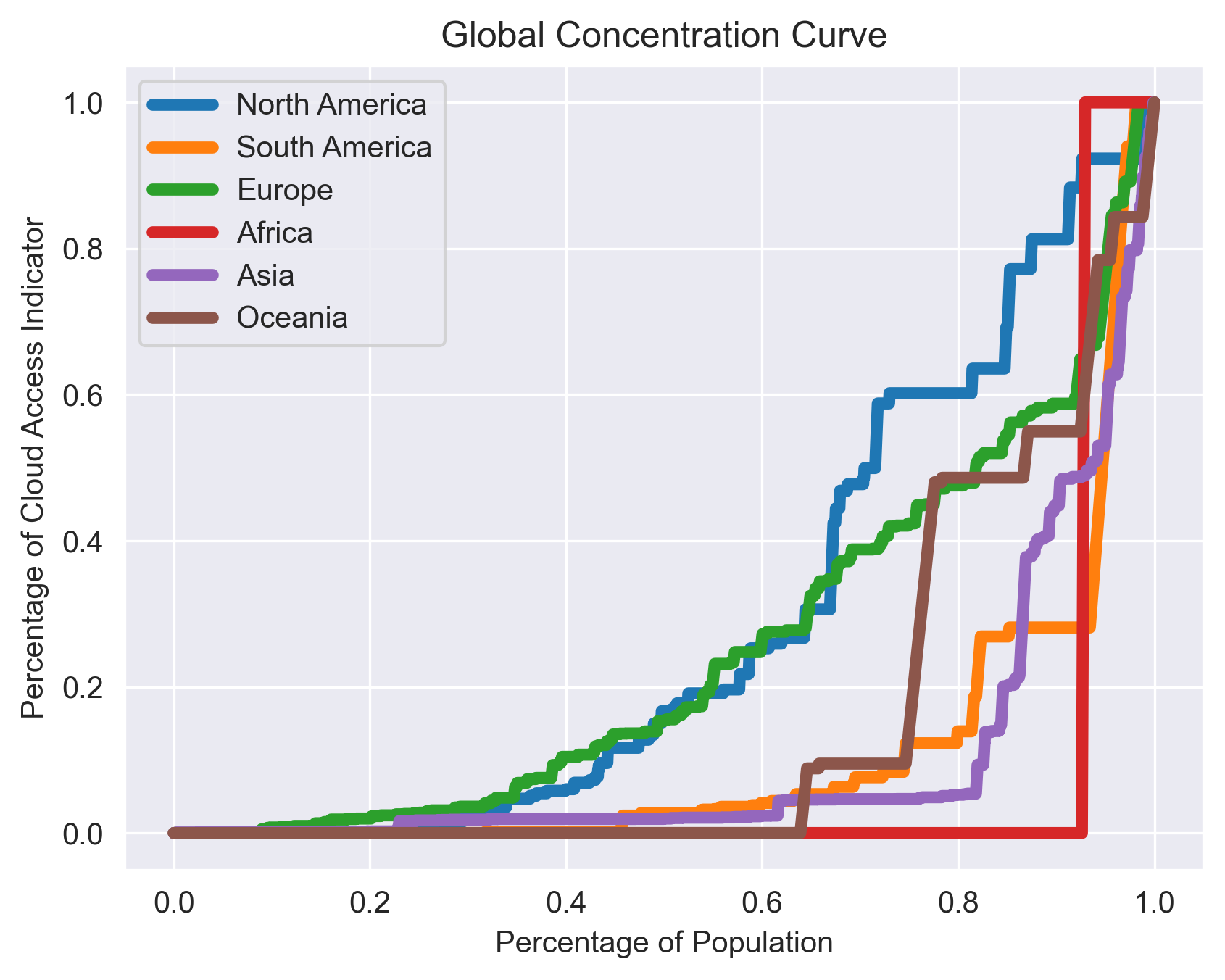}\label{fig:region_concentration}}
    \subfloat[Concentration Curves - Edge]
    {\includegraphics[width=0.32\textwidth]{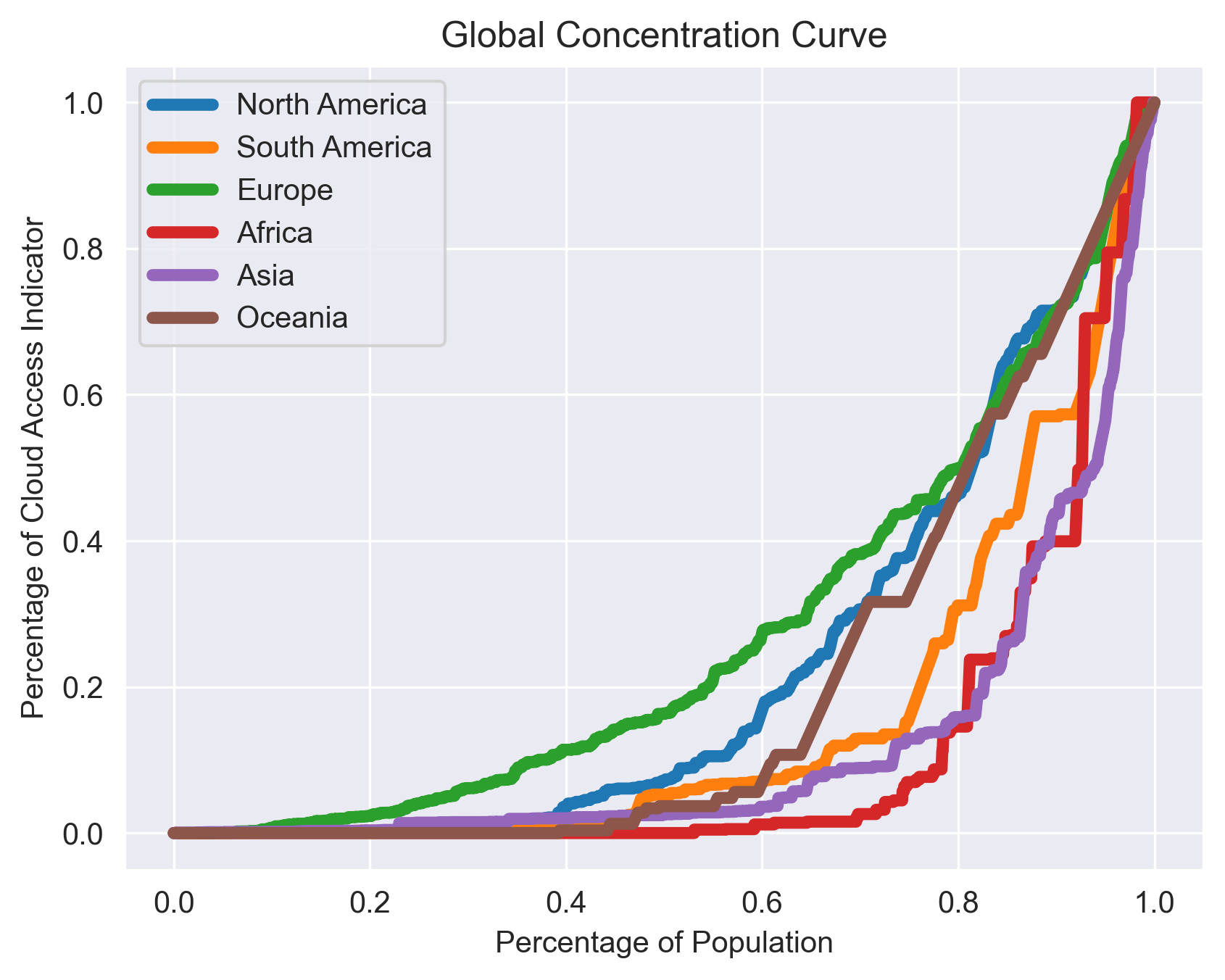}\label{fig:edge_concentration}}
    \subfloat[Change in CI as new Local Zones are launched]
    {\includegraphics[width=0.32\textwidth]{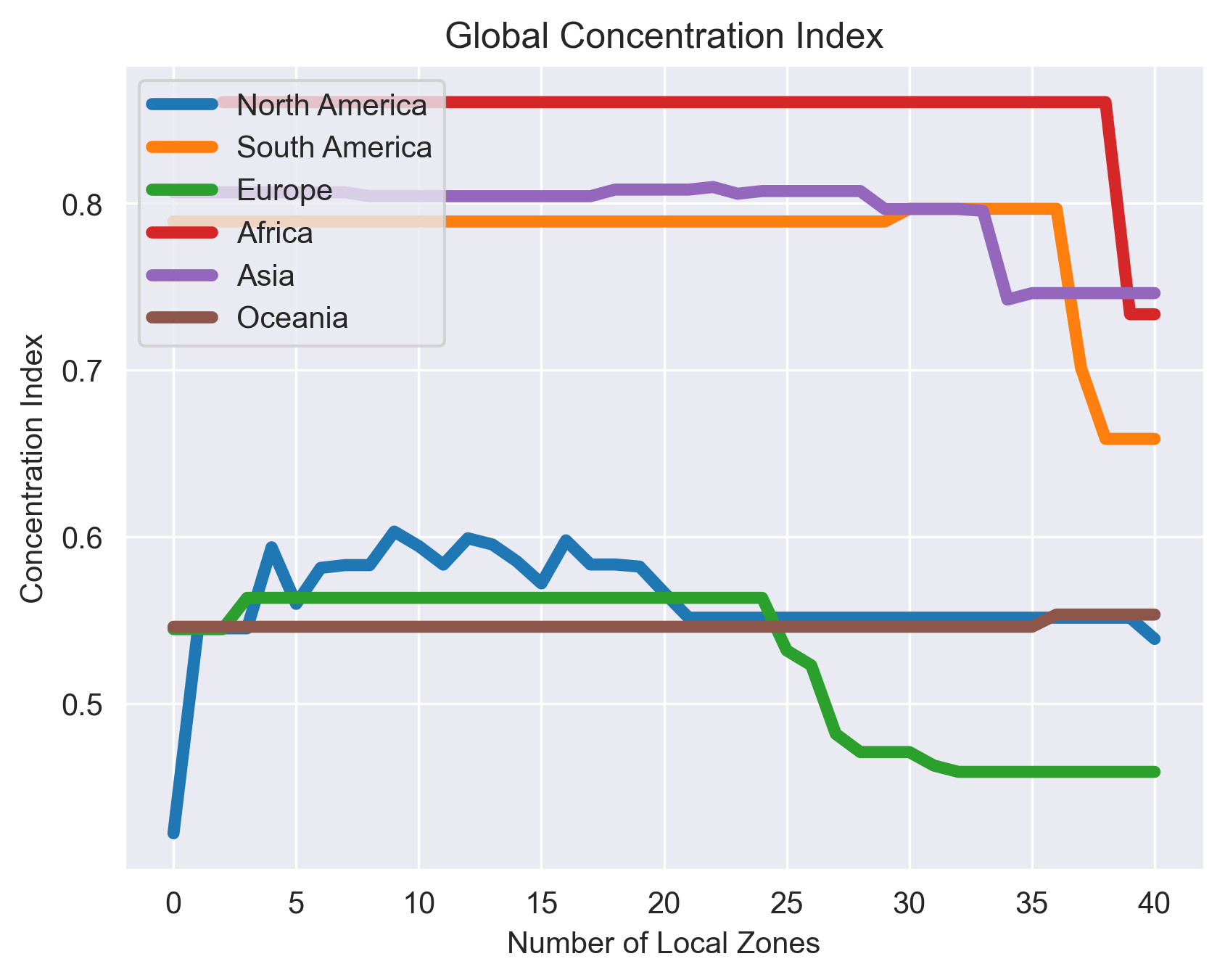}\label{fig:ci_new_locations}}
    \caption{Concentration curves globally}
\end{figure}

At the continent level we again see unfairness favoring higher income areas because the concentration curves are below the line of equality. Fig~\ref{fig:region_concentration} shows these concentration curves for each continent when only considering AWS regions. Africa has the highest CI, 0.86, due to the only datacenter region on the continent being in an area within the 90th percentile of NTL. Fig~\ref{fig:edge_concentration} expanded the datacenters considered to include all edge locations, including Local Zones. Due to the 3 additional locations in Africa at the network edge, the CI drops to 0.79. This is similar to what we saw when Local Zones were added in the US.

However, there isn't always an improvement to the concentration index as new datacenter locations are added. North America in particular, which has the lowest CI of all continents when only considering regions, has a higher CI with Local Zones. This can be explained by Fig~\ref{fig:ci_new_locations}, which shows the change in CI per-continent as new datacenters are added, starting with the first Local Zone in Los Angeles. This first new datacenter in Los Angeles causes the jump in CI since many people in a high NTL area are within the threshold, $\sigma$, to a datacenter. Overall, we see similar global trends with NTL as we saw in the US with income data from the ACS - new locations tend to decrease the CI, but the unfairness is still high. The largest decrease is observed in Africa with the most recent launch of a datacenter in Lagos, Nigeria.

The changes in CI with each launch also demonstrates the per-region rollout that new locations have been following. The first half of new locations are primarily in the US. Not until all the US locations are launched do we see Europe start to gain new locations, and only after these launched we see significant activity in the rest of the world. There is a large gap between the CI values of continents with more developed networking infrastructure (North America, Europe, and Oceania) and the continents that have historically received lower investments in networking infrastructure (Asia, Africa and South America). Globally, unfairness is still very high. Since the first Local Zone it has reached a maximum of 0.77, and is now only slightly lower at 0.70. While countries or continents with more investments from cloud providers are getting more fair, there are still high levels of unfairness globally.

\subsubsection{Optimal selection}\label{sec:optimal_selection}

\begin{wrapfigure}{r}{0.50\textwidth}
  \begin{center}
    \includegraphics[width=0.48\textwidth]{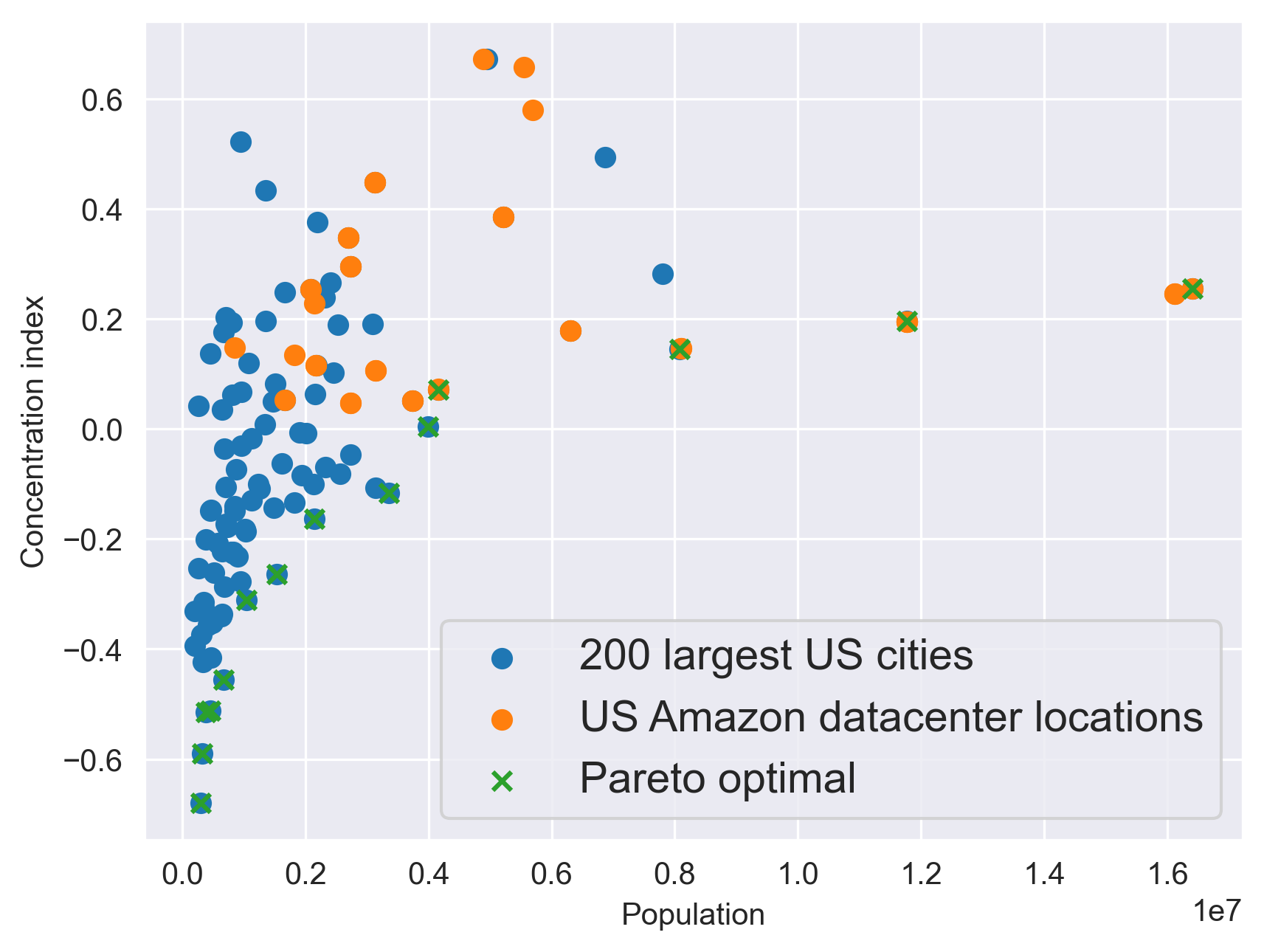}
  \end{center}
  \caption{Tradeoff between population and concentration index for the largest US cities and AWS locations. Pareto optimal cities are marked.}
  \label{fig:optimal_cities}
\end{wrapfigure}

Cloud providers may be more interested in increasing coverage of the cloud edge by targeting cities with the highest population rather than having a goal to minimize unfairness. Next, we demonstrate a strategy to optimize for both. With our framework for evaluating fairness, we treat this as a multi-objective optimization problem to balance the fairness/coverage tradeoff.  We computed total population within $\sigma = 70km$ and the CI value for each of the top 200 most populated US cities (excluding cities that are less than $\sigma$ away from any previously considered city), using the same methods as \S\ref{sec:us_unfairness}. From this, the pareto-optimal cities are ones that lie on the front of high population but low concentration index. While a negative CI is still unfair (0 is maximum fairness) the cities with a negative CI are needed to balance the CI from the high values of the largest population cities. Fig~\ref{fig:optimal_cities} demonstrates this pareto-optimal selection of cities, as well as the ones actually used for Local Zones.

For the most part, the largest population cities all have datacenters. The only two cities that are high population but not current datacenter locations are Washington D.C. and Riverside, CA. Both of these are above our 70km threshold to other cities, but are still fairly close to existing datacenters in Los Angeles and North Virginia which can explain why they don't have their own dedicated datacenter. All the AWS locations have a CI > 0, again demonstrating an unfairness that gives higher income communities closer access to datacenters. There are 4 cities on the pareto front that cover more population than some existing locations, but do not currently have an AWS datacenter: El Paso, Detroit, Memphis, and Tampa. Each of these cities could have been selected in place of a less optimal city to provide greater than or equal population coverage while being more fair.




\section{Case Study LEO Satellite ISP}\label{sec:leo_satellites}

Satellite internet presents a solution for providing Internet access where it is not practical to deploy traditional fiber networks. They are already being used particularly in African countries to offer access to many users at once in public places~\cite{satellite_all}. Due to their increasing popularity in underserved regions, a discussion of the cloud's digital divide would not be complete without looking at it from the perspective of satellite ISPs. In this section, we focus specifically on a new class of satellite internet - LEO satellites.

\begin{figure}
    \centering
    \subfloat[width=0.40\columnwidth][Starlink latency. Dashed lines are the RTT to the NAT hop, which represents the latency to reach the ground station. The ground station latency for both regions and the edge is similar to the total RTT for the edge.]
    {\includegraphics[width=0.40\columnwidth]{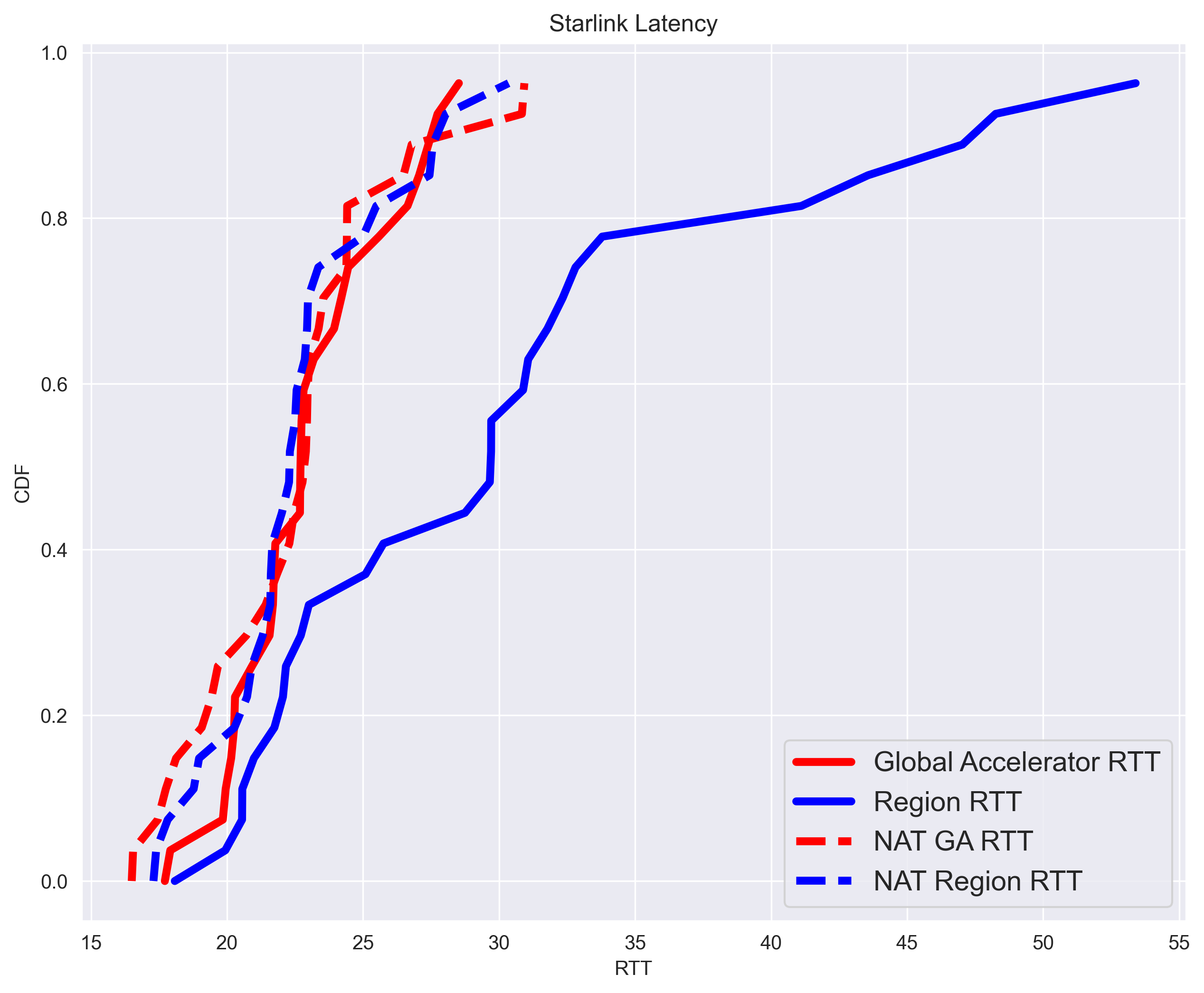}\label{fig:starlink_cdfs}}
    \subfloat[width=0.40\columnwidth][CI for larger values of $\sigma$.]
    {\includegraphics[width=0.40\columnwidth]{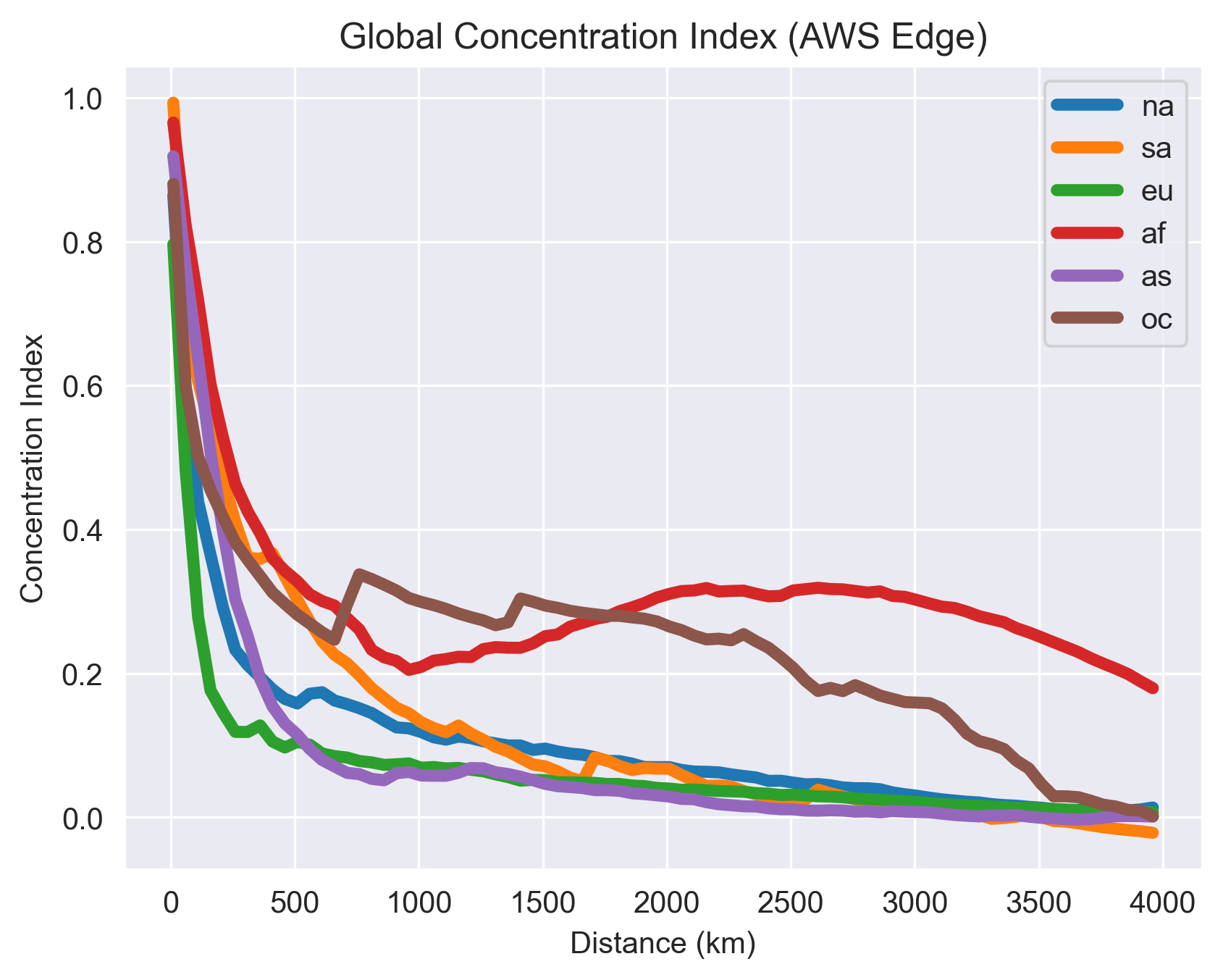}\label{fig:leo_unfairness}}

    \subfloat[width=0.80\columnwidth][Percentile ratios for LEO distances]
    {\includegraphics[width=0.80\columnwidth]{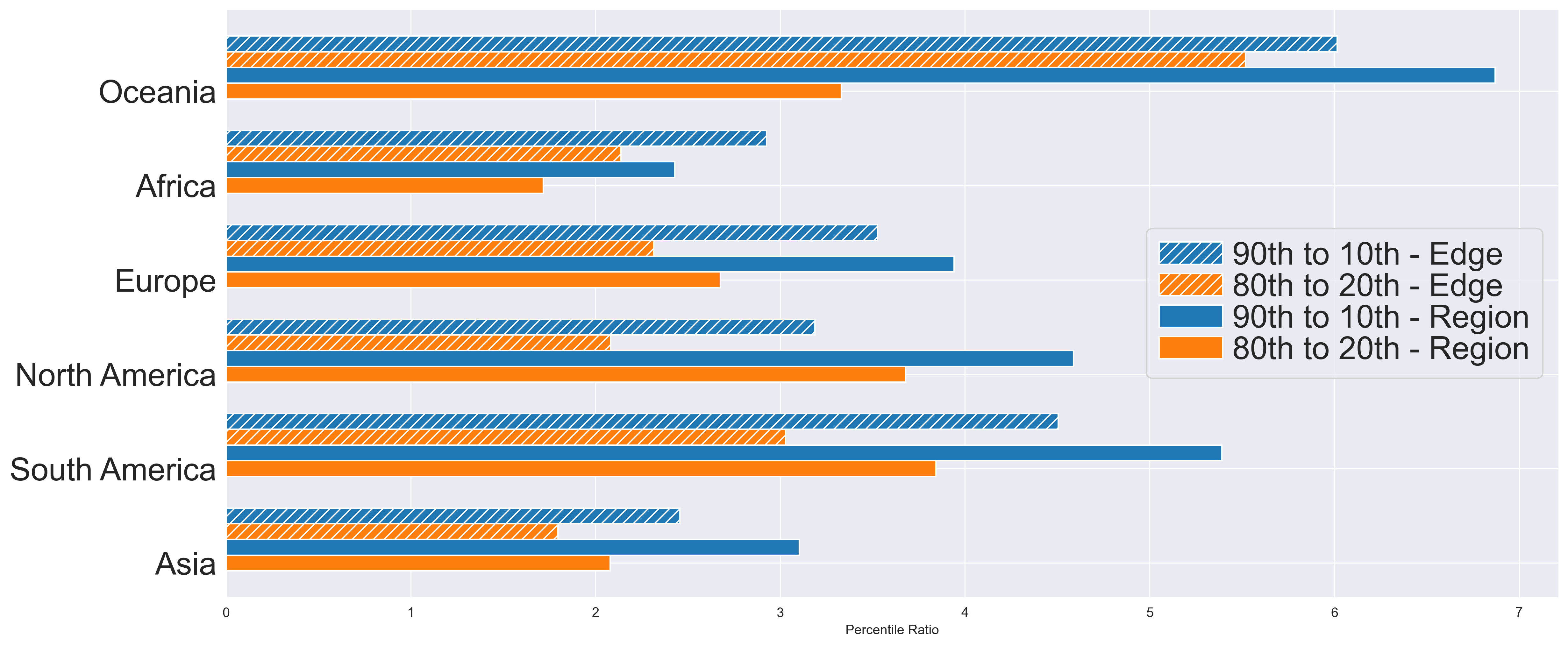}\label{fig:leo_inequality}}
    \caption{Case study of results using LEO satellite ISPs}
\end{figure}

Internet from LEO satellite constellations can already provide low-latency access to the cloud edge. The RTT to reach Amazon's edge network using Global Accelerator is comparable to the latency spent until the end of the satellite hop - when packets reach the ground station. This result is shown in Fig~\ref{fig:starlink_cdfs} which plots CDFs of RTT to the ground stations and endpoints of Amazon's edge network and regions. Latency to regions are considerably longer which is due to time spent in the private WAN. This tells us satellite ground stations are geographically close and well connected to cloud edge datacenters, but can be far from regional datacenters.

Our measurements consisted of 9k traceroutes from all 42 RIPE Atlas probes on the Starlink network (identified using ASN) targeting the Global Accelerator IP address and VMs in AWS regions used in our earlier experiments. Starlink uses a carrier-grade network address translation (NAT) at the exit of the satellite link~\cite{first_look_starlink}, which allows us to identify the satellite portion of the latency as the $100.64.0.1$ hop in the traceroute. These probes are only in North America, Europe, and Oceania - the same continents with the best performing network infrastructure from our previous measurements. At the time of our experiments there was limited deployment in other continents, with coverage expected to expand as more ground stations are built.

While minimum RTT is currently above the 20ms threshold for more than half of the probes, it is expected to drop. Due to the satellites' low orbit, speed of light constraints limit the RTT to only 4ms. In practice latency will be longer, but is expected to be reduced to under 10ms~\cite{starlinklatency}. This is low enough that LEO satellites are a promising option to support new low latency applications running on edge networks.

The minimum RTT we observed is very consistent, with ~10ms difference between the fastest and slowest probe. This is because the paths are the same for almost all probes, regardless of location. Each path consists of the local network to satellite dish, satellite hop, and the ground station to AWS. Ideally the local network has minimal latency, and we observe the ground stations to be close to cloud edge networks. Based on the traceroute IP path, 80\% of our probes could reach Amazon's edge without using another Autonomous System, confirming the homogeneity of this network across different geographies.

This similarity guides our intuition that LEO satellite networks can provide equal and fair access to all, which we verify through the percentile ratio and concentration index. Fig~\ref{fig:leo_inequality} shows the percentile index measurements from \S\ref{sec:world_inequality} with an extra 500km distance added to approximate the distance travelled in the satellite hop. This outweighs other differences in distance, and greatly equalizes access. In many continents, the expansion of cloud edge computing actually reduces the inequality for this scenario, in contrast to the the previous evaluation where it always increased.

Satellites can also reach a much larger area with low delay. A single satellite covers a ground area with a radius of ~900km~\cite{without_isl}. Furthermore, while current networks do not use inter-satellite links, when these are enabled latency is expected to be even lower due to direct line of sight between satellites and the speed of light in space being faster than in fiber~\cite{without_isl}. In this case, a single edge datacenter can provide low latency access to a wider area, changing our choice of $\sigma = 70km$ in \S\ref{sec:unfairness}. Fig~\ref{fig:leo_unfairness} plots the CI as $\sigma$ is increased. In general, the value approaches 0 (perfect fairness) the more we can increase $\sigma$. An ideal speed of light internet can reach 3000km with a 20ms RTT, but even 1/3 of that results in much more fair access to cloud compute.

\section{Related Work}\label{sec:related_work}

\subsection{Cloud and Edge Latency}

Many previous measurement studies of cloud access latency have focused on traditional cloud regions, not the cloud edge. Corneo et al. used RIPE Atlas probes to measure the global latency to the cloud~\cite{surrounded}. They found that North America, Europe, and Oceania have sufficient datacenter infrastructure to support many emerging applications, but that other continents would require additional datacenters. Dang et al. performed a similar study using SpeedChecker \cite{speedcheker}, which demonstrated much higher latency to the cloud when using wireless probes (Wi-Fi and LTE)~\cite{cloudy} . Both these studies considered many cloud providers with and without a private WAN, but neither included datacenters at the edge of cloud private WAN.

Recent work has also measured performance characteristics of networks between public clouds. Rotman et al. measure latency between the same three cloud providers as our measurements, but only including cloud regions~\cite{cloudcast}. The Skyplane system~\cite{skyplane} optimizes large transfers between cloud regions by measuring the highest bandwidth path. In both cases, these measurements are for inter-cloud links, while we focus on the edge connecting users to cloud networks.

Measurements of edge compute have been restricted to hypothetical deployments or focused on one region. Corneo et al. use traceroutes from RIPE Atlas probes to cloud datacenters in the US to identify routers best suited to have nearby edge locations~\cite{howMuch}. They find latency improvements in the US of up to 30\% are possible from deploying edge computing, while the absolute value of this improvement would be ~3ms. Previous measurement studies have also targetted 6k+ Akamai edge servers to compare these with cloud datacenters around the world~\cite{latencyComparison}. These CDN edge servers are much more prevalent than the cloud edge that our measurements target.  Additionally, prior work has compared Alibaba's edge compute service in China to their cloud compute service based on a number of factors including latency and application QoE~\cite{firstLook}. To the best of our knowledge, there have not been prior measurement studies on global latency to commercially available cloud edge.

\subsection{Internet Access Inequality}

Prior work has revealed aspects of the digital divide both in the US and around the world, particularly in developing regions. The availability, or lack thereof, of ISPs has been studied at the level of US counties~\cite{technoEconomicBroadband}. Additionally, examining Internet providers in LA has shown that low income and minority census blocks have less access to broadband upgrades - demonstrating the inequality in Internet access is unfairly leaving these communities behind~\cite{whoGetsAccessLA}. Paul et al. conducted a study across California using crowdsourced speed measurements to show how several demographic attributes such as income and education relate to internet quality, not just access~\cite{characterizingCaliforniaInternetQuality}. Previous work has also surveyed digital literacy in developing countries, which contributes to the digital divide as well~\cite{digitalLiteracy}. In our study, we use global demographic data to characterize the unfairness in datacenter locations - which determines the minimum possible latency for cloud applications and introduces this metric as a new aspect affecting the digital divide.
Finally, others work in the context of developing regions have looked at performance implications of free services~\cite{missit, freebasics-ccr}.

\section{Discussion}

\textbf{Compute Inside Satellite Networks.} Our work demonstrates the selection of cities for cloud edge datacenters creates a digital divide where some communities have advantages over others. As new applications are developed we expect the consequences of this divide to result in unfair access to new technologies that support health, education, or economic activity. This observation may be used to motivate deployments of new networks such as satellite ISPs, as was highlighted in the case study. These networks help level the playing field but the resulting latency (which is similar for all users) may be too high for certain applications. For example, there are already specialized applications such as high frequency trading where sub-millisecond latency matters~\cite{birds_eye}. For these applications just making the network faster still results in inequalities between those close and far from a datacenter. However, there have been other proposed systems such as allowing compute to move rather than be fixed in one place~\cite{in_orbit_compute} which we expect to help reduce inequalities.

\textbf{Other Considerations in Datacenter Selection.} There are also other aspects to datacenter locations that affect the digital divide which we plan to consider in future work. For instance, data sovereignty laws including GDPR may require some people to use applications running in select countries. In these situations not all datacenters are equally useful, and it may severely limit availability of edge compute. There are also sustainability concerns when selecting a datacenter region. The electric grid serving each datacenter will have its own carbon intensity - the amount of CO2 released by that grid. Further influencing this is the location of clean energy projects created by cloud providers to lower the emissions of their datacenters. Providing cloud services without exceeding emission targets will further limit available datacenter locations, and may make it easier for some areas to decarbonize.

\section{Conclusion}
Our study quantifies the latency improvements currently possible and expected to be possible with the cloud edge. The cloud edge is not serving everyone equally, and widens the digital divide between those with the closest and furthest datacenters. Furthermore, we measure the locations selected for cloud edge datacenters to demonstrate unfairness that gives wealthier communities more access. Using datasets from the ACS and NTL satellite images, we demonstrated the high concentration of datacenters in high income communities, even while cities with higher populations are lacking datacenters. We believe our work looks at the digital divide from a new and important perspective -- latency to the nearest cloud location --  and provides guidance on promising technologies and deployment paths that could help in reducing the digital divide.

\section*{Acknowledgements}
Thanks to the Tufts NAT Lab for their feedback and support, and to the RIPE Atlas community for generous donations of credits.

\noindent This work was partially funded by NSF CNS award: 2106797. 
\newpage
\bibliography{main}
\bibliographystyle{abbrvnat}

\end{document}